\documentclass[preprint,prc,tightenlines,showpacs,superscriptaddress,nofootinbib]{revtex4-1}
\usepackage{amsfonts}
\usepackage{amssymb}
\usepackage{amsmath}
\usepackage{graphicx,color}
\usepackage{dcolumn}
\usepackage{epsfig}
\usepackage{bm}
\usepackage{bbm}
\usepackage{ulem}

\usepackage{color}

\newcommand{\unittwo}{\mathbbm{1}}

\newcommand{\ket}[1]{|{#1}\rangle}
\newcommand{\bra}[1]{\langle{#1}|}
\newcommand{\inp}[2]{\langle{#1}|{#2}\rangle}

\def\gtap{\ \raise.3ex\hbox{$>$\kern-.75em\lower1ex\hbox{$\sim$}}\ }
\def\ltap{\ \raise.3ex\hbox{$<$\kern-.75em\lower1ex\hbox{$\sim$}}\ }
\begin{document}
\begin{flushright}
 LFTC-18-6/27, J-PARC TH-0129
\end{flushright}

\title{
   Nuclear applications of ANL-Osaka amplitudes: pion  
     photo-productions on deuteron
}
\author{S. X. Nakamura}
\affiliation{
Laborat\'orio de F\'isica Te\'orica e Computacional - LFTC, 
Universidade Cruzeiro do Sul, S\~ao Paulo, SP 01506-000, Brazil
}
\author{H. Kamano}
\affiliation{Research Center for Nuclear Physics, Osaka University, Ibaraki, Osaka
567-0047, Japan}
\author{T.-S. H. Lee}
\affiliation{Physics Division, Argonne National Laboratory, Argonne, Illinois 60439, USA}
\author{T. Sato}
\affiliation{Research Center for Nuclear Physics, Osaka University, Ibaraki, Osaka
567-0047, Japan}
\affiliation{J-PARC Branch, KEK Theory Center, IPNS, KEK, Tokai, Ibaraki 319-1106, Japan}

\begin{abstract}
 The  Argonne National Laboratory-Osaka University
      (ANL-Osaka) amplitudes 
      (http://www.phy.anl.gov/theory/research/anl-osaka-pwa/)
       are applied to study  pion  photo-production reactions  on the deuteron target. 
Within the multiple scattering
formulation, we predict the cross sections of $d(\gamma,\pi)NN$ in the nucleon resonance region.
The calculations include the impulse term and the final-state
interaction (FSI) terms due to pion-exchange  and nucleon-exchange. 
We show that
the off-shell effects, calculated from the meson-exchange mechanisms, on the propagations of the
exchanged nucleon and pion  are significant in determining the reaction amplitudes.
The FSI effects on the predicted cross sections
 are found to be important at energies near the $\Delta$(1232) resonance,
and are still significant at higher energies.
The results are in good agreement with most of the available data of $d(\gamma,\pi^0)np$ and
$d(\gamma,\pi^-)pp$ reactions.
\end{abstract}


\maketitle

\section{Introduction}
\label{intro}
The developments of dynamical models for $\pi N$ and $\gamma N$ reactions were motivated
by the success of the meson-exchange models~\cite{machleidt} for the $NN$ interactions and the 
earlier attempts~\cite{donna}
to relate the isobar models
 of nucleon resonances to the predictions of constituent quark models~\cite{capstick}.
Within the formulation given in Refs.~\cite{sl96,msl07}, the development of  
dynamical models of $\pi N$ and $\gamma N$
reactions has two main objectives:
\begin{enumerate}
\item Develop interpretations of the nucleon resonances ($N^*$) within
the framework that the excitations of the  nucleon can be described
in terms of  bare baryon states and  meson-exchange interactions.
 With the parameters
determined by fitting the data of $\pi N$ and $\gamma N$ reactions up to
the invariant mass $W=2$~GeV, 
the poles and residues of  nucleon
resonances are extracted by performing analytic continuations
 of the partial-wave amplitudes predicted within the constructed model
to the complex energy-plane.
\item Apply the constructed dynamical model to investigate the production and propagation of
mesons and nucleon resonances in nuclei, which are crucial for analyzing
data from experiments on nuclear targets in the nucleon resonance region, such as
the recent experiments on the neutrino oscillations~\cite{nu-exp}.
\end{enumerate}

With the efforts reported in 
Refs.~\cite{jlms07,jlmss08,djlss08,ssl09,kjlms09a,jklmts09,kjlms09b,sjklms10,knls10,ssl10,shkl11,knls13,knls16},
we have achieved the first objective,
 and the resulting  
Argonne National Laboratory-Osaka University
(ANL-Osaka) amplitudes has now become available~\cite{web} to public.
In this paper, we will take a  step toward reaching  the second objective 
by applying  the constructed dynamical coupled-channel model (ANL-Osaka model)
 presented in Ref.~\cite{knls13,knls16} to investigate pion photo-production on the simplest
nucleus, the deuteron ($d$). We will focus on the
$d(\gamma,\pi)NN$ reactions for the following reason.
An uncertainty in the construction of the ANL-Osaka model was in the determination of
the isospin structure of the nucleon resonances by fitting the available data
of $\gamma n \rightarrow \pi^- p$ which were extracted from the $d(\gamma,\pi^-)pp$ data with
some procedures to select the  $\gamma n\rightarrow \pi^-p$ events. 
Furthermore, $\gamma n\rightarrow \pi^0n$ data included in our previous fit were
      rather scarce.\footnote{
For determining the isospin structure of nucleon resonances,
$\gamma p \rightarrow \pi^0 p, \pi^+n$ data and 
either of $\gamma n \rightarrow \pi^- p$ or $\gamma n\rightarrow \pi^0n$ data are
needed in the fit if the data are of very high accuracy.
This is however not the case, thus the fits including both of the neutron data may
be needed to reduce the uncertainty.}
 We therefore need to examine whether our predictions of $d(\gamma,\pi)NN$
will be in agreement with the original data of $d(\gamma,\pi^-)pp$ and $d(\gamma,\pi^0)pn$.

One of the important features of the ANL-Osaka 
model is that the amplitudes are generated from an energy $independent$ Hamiltonian
which can be included in  the conventional Hamiltonian formulation
 for developing many-body descriptions of nuclear reactions. 
This is achieved by using a unitary transformation 
method~\cite{ut-osaka,sl96} 
to derive hadron-hadron
interactions from relativistic quantum field theory with meson and baryon degrees of freedom.
This feature allows us to use rigorously 
the well-established multiple-scattering 
formulation~\cite{feshbach}
to develop a reaction model for $d(\gamma,\pi)NN$ reactions. For our limited purpose here,
we follow the previous 
investigations~\cite{arenhover,fix,lev06,sch10,wsl15,tara,tara-1}
to only include the single scattering (impulse) amplitudes and the
 double scattering amplitudes  due to pion-exchange and nucleon-exchange mechanisms.
 The exchanges of unstable
particles $\rho$, $\sigma$
and $\Delta$(1232), which can be generated from the ANL-Osaka model, are more difficult to
calculate and are  neglected in this work for simplicity. 

Thus our task is to develop formula for calculating
the amplitudes illustrated in Fig.~\ref{fig:diag}. The $\gamma N\rightarrow \pi N$ 
and  $\pi N\rightarrow \pi N$ amplitudes  for evaluating the pion
re-scattering term of Fig.~\ref{fig:diag}(c) 
can be  generated from the ANL-Osaka model. The initial deuteron wave function and 
the $NN\rightarrow NN$
amplitudes in Fig.~\ref{fig:diag}(b) can be generated from any of the available 
high-precision $NN$ potentials. 
To be consistent with the meson-exchange
mechanisms of the ANL-Osaka model, we choose
the CD-Bonn potential~\cite{cdbonn}.
 There are two important issues  
in practical calculations of the matrix elements of these mechanisms, as discussed well in
the earlier investigations~\cite{thomas,lee75} of the multiple scattering of
hadrons  from nuclei within
the Hamiltonian formulation.
First, we need to define a Lorentz boost transformation to relate
the $\gamma N \rightarrow \pi N$, $\pi N\rightarrow \pi N$ and $NN\rightarrow NN$ amplitudes
in Fig.~\ref{fig:diag} in the photon-deuteron laboratory frame to those
in the two-body center of mass (CM)
frame where the two-body amplitudes are generated from Hamiltonian by solving
scattering equations in partial-wave representation. 
In particular, the spin rotations must be taken 
into account relativistically for investigating polarization observables.
Here we follow the method of relativistic quantum mechanics, as detailed
in Ref.~\cite{polyzou,polyzou-1,kei-pol}. The second issue is that 
the resulting FSI amplitudes will include loop-integrations over
the off-energy-shell matrix elements of the two-body amplitudes in Fig.~\ref{fig:diag}. Thus it is necessary to 
introduce an approximation to choose the collision energies in generating these
 two-body off-shell matrix elements.
Here we will use an approach which accounts for the energy shared by the spectator 
nucleon and pion in Fig.~\ref{fig:diag}.

The organization of the rest of this paper is as follows.
In Sec.~\ref{sec:model}, we present formulas for calculating the amplitudes
of the impulse term [Fig.~\ref{fig:diag}.(a)], 
nucleon re-scattering ($N$-exchange) term [Fig.~\ref{fig:diag}.(b)],
 and pion re-scattering ($\pi$-exchange) term [Fig.~\ref{fig:diag}.(c)].
The results for comparing our predictions with 
 the available data of $d(\gamma,\pi^0)pn$ and
$d(\gamma,\pi^0)pp$, and  for examining the importance of FSI and
off-shell effects will be given in Sec.~\ref{sec:result}. 
A summary is given in Sec.~\ref{sec:summary}.

\section{Formulation}
\label{sec:model}
In this section, we describe the theoretical formulation used in our investigation of
$d(\gamma,\pi)NN$ reactions. All of the formulas for calculations are presented with 
 the normalizations: $\langle\bm{k}'|\bm{k}\rangle=\delta^{(3)}(\bm{k}'-\bm{k})$
for plane-wave states and $\langle\Phi_d|\Phi_d\rangle=1$ for bound states.

\subsection{Model Hamiltonian with meson and baryon  degrees of freedom}
Our starting point is 
the following
many-body Hamiltonian
\begin{eqnarray}
H= H_0 + \frac{1}{2}\sum_{i\neq j} V_{NN,NN}(i,j)+ \sum_{i}h^I_{\rm AO}(i) \ ,
\label{eq:tot-h}
\end{eqnarray}
where $H_0$ is the free Hamiltonian for all particles in the considered processes, 
$V_{NN,NN}(i,j)$ is a 
 nucleon-nucleon potential between the nucleons $i$ and $j$, and
\begin{eqnarray}
h^I_{\rm AO}(i)= \sum_{c,c'}v_{c,c'}(i) +
\sum_{N^*}\sum_{c}[\Gamma_{N^*,c}(i)+\Gamma^\dagger_{N^*,c}(i)] \ ,
\label{ao-h}
\end{eqnarray}
is the interaction Hamiltonian of the ANL-Osaka (AO) model.
The channels included are
$c,c'=\gamma N, \pi N, \eta N, K\Lambda, K\Sigma$ and $\pi\pi N$ with resonant
$\pi\Delta,\rho N$, and $ \sigma N$ components. The energy independent meson-exchange 
potentials $v_{c,c'}$ are derived from 
phenomenological Lagrangians by using  the unitary transformation method~\cite{sl96,ut-osaka}.
The vertex interaction  $\Gamma_{N^*,c}$ defines 
the formation of a bare $N^*$ state 
from a channel $c$.
The parameters of the Hamiltonian $h^I_{\rm AO}$ have been determined in Refs.~\cite{knls13,knls16}
by fitting about 26,000 data points
of the $\pi N$ and $\gamma N$ reactions up to the invariant mass
$W\simeq 2$~GeV.

Here we mention that the model Hamiltonian defined above is a very significant
extension of the Hamiltonian developed in Refs.~\cite{lee83,mitzutani} 
where only $\pi$, $N$, and $\Delta$ degrees of freedom are included.
It can be used to investigate 
$\pi$, $\eta$, and two-pion production reactions on nuclei in the nucleon resonance region.
The first attempt in this direction was made to investigate
the $d(\gamma,\eta)pn$ reaction in the
context of extracting the low-energy $\eta N$ scattering parameters~\cite{etaN}.
Here we focus on the pion photo-production on the deuteron to test our predictions.  

\subsection{Scattering amplitudes for $d(\gamma,\pi)NN$ reactions}

\begin{figure}[t] \vspace{-0.cm}
\begin{center}
\includegraphics[width=0.8\columnwidth]{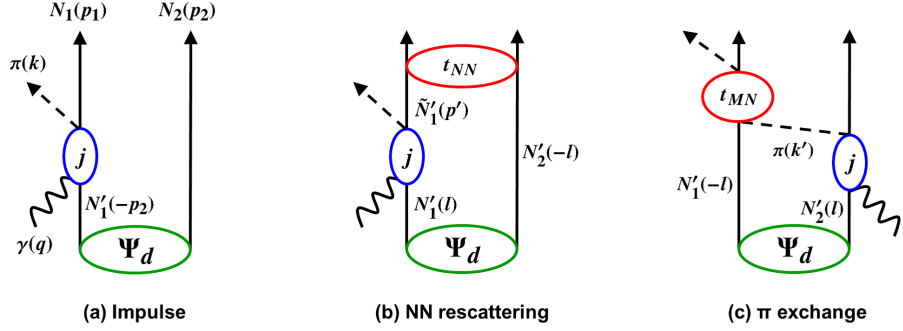}
\caption{
Diagrammatic representation of reaction mechanisms considered in this work for $d(\gamma,\pi)NN$:
(left) impulse, (center) $NN$ rescattering ($N$-exchange),
(right) $\pi N$ rescattering ($\pi$-exchange) mechanisms.
}
\label{fig:diag}
\end{center}
\end{figure}

Starting with the scattering $T$-matrix given by
the Hamiltonian Eq.~(\ref{eq:tot-h}), 
it is straightforward to follow the well-developed procedures~\cite{feshbach}
to obtain  a multiple scattering formulation
of the $d(\gamma,\pi)NN$ reactions. Following the previous 
investigations~\cite{arenhover, wsl15, tara, tara-1},
we will only keep
 the single-scattering (impulse) terms and the double scattering terms with $\pi NN$ 
intermediate states.
The $T$-matrix operator for the $d(\gamma,\pi)NN$ reactions
can then be written as
\begin{eqnarray}
T(E)&=& \sum_{i} \tau_{\pi N_i,\gamma N_i}(E)+
\sum_{i\neq j } \tau_{N_iN_j,N_iN_j}(E)
\frac{|\pi N_iN_j\rangle \langle\pi N_iN_j|}{E-H_0+i\epsilon}
\tau_{\pi N_i,\gamma N_i}(E)\nonumber \\
&&
+
\sum_{i\neq j } \tau_{\pi N_j,\pi  N_j}(E)
\frac{|\pi N_iN_j\rangle \langle\pi N_iN_j|}{E-H_0+i\epsilon}
\tau_{\pi N_i,\gamma N_i}(E) \ ,
\label{eq:amp0}
\end{eqnarray}
where $E$ is the total energy of the $\gamma d$ system,
$\tau_{\alpha N_i,\beta N_i}(E)$ is a scattering operator associated with the
$i$-th nucleon in the deuteron. Note that the matrix elements
of $\tau_{\alpha N_i,\beta N_i}(E)$ are determined by the total Hamiltonian 
Eq.~(\ref{eq:tot-h}) and cannot be calculated exactly.
Therefore, we adopt the spectator
approximation~\cite{thomas,lee75} as
\begin{eqnarray}
\tau_{\alpha,N_i,\beta N_i}(E) \sim t_{\alpha,N_i,\beta N_i}(E-E_{\rm spectator}) \ ,
\label{eq:spect}
\end{eqnarray}
where $t_{\alpha,N_i,\beta N_i}(E')$ is the two-body scattering operator in free space,
and $E_{\rm spectator}$ is the energy of the spectator of the two-body scattering  
in the $\gamma NN$ or $\pi NN$ three-particle states, as seen in Fig.~\ref{fig:diag}.
The resulting $T$-matrix operator is consistent with that of the Faddeev
framework up to and including the double scattering terms, 
as has been also discussed in Ref.~\cite{etaN}.

To proceed further, we need to define the matrix elements of the scattering operator $t_{c,c'}(W)$
with an  invariant mass $W$.
Within the ANL-Osaka model, these
matrix elements are generated 
in the two-body CM frame
by solving the following 
coupled-channel equations in each partial wave:
\begin{eqnarray}
\langle k|t_{c,c'}(W)|k'\rangle&=&\langle k| V_{c,c'}(W)|k'\rangle\nonumber \\
&& +\sum_{c''}\int k^{''\,2} dk^{''}
\frac{\langle k|V_{c,c''}(W)|k''\rangle\langle k''|t_{c'',c'}(W)|k'\rangle}
{W-E_{c'',1}(k'')-E_{c'',2}(k'')-\Sigma_{c''}(k'',W)+i\epsilon} \ ,
\label{eq:cceq}
\end{eqnarray}
where $k$, $k'$ and $k''$ are  the momenta in the CM frame, 
$E_{c,i}(k)=\sqrt{m_{c,i}^2+k^2}$ is the energy of a particle $i$ in a channel $c$
with the mass $m_{c,i}$,
$c,c',c''=\pi N, \eta N, \pi\Delta,\rho N, \sigma N, K\Lambda, K\Sigma$
are the considered channels, and $c'=\gamma N$ channel is included
perturbatively;
$\Sigma_{c''}$ is the self-energy for the unstable channels
$c''=\pi\Delta,\rho N, \sigma N$, and is zero for the other stable channels.
The driving term is
\begin{eqnarray}
V_{c,c'}(W)=v_{c,c'}+\sum_{N^*}\Gamma^\dagger_{N^*,c}\frac{1}{W-M_{N^*}}\Gamma_{N^*,c'}
+ Z(W)\ ,
\end{eqnarray}
where $M_{N^*}$ is the bare mass of an excited nucleon state $N^*$;
$Z(W)$ is a particle-exchange Z-diagram in which a $\pi\pi N$ channel is included.
Note that the initial and final states of each matrix element in Eq.~(\ref{eq:cceq}) can be
on-energy-shell  [$W=E_{c,1}(k)+E_{c,2}(k)=E_{c',1}(k')+E_{c',2}(k')$] or 
off-energy-shell [$W\ne E_{c,1}(k)+E_{c,2}(k)\neq E_{c',1}(k')+E_{c',2}(k')$] within the Hamiltonian
formulation.

By using Eqs.~(\ref{eq:amp0}) and (\ref{eq:spect}) and
the momenta defined in Fig.~\ref{fig:diag}, the Lorentz 
invariant scattering amplitude
of $d(\gamma,\pi)NN$  can then be written as
\begin{eqnarray}
M_{f,i}(E) &=&
\sqrt{  { (2  E_\pi(\bm{k}))
E_N(\bm{p}_1)E_N(\bm{p}_2)}\over m_N^2}\nonumber \\
&&\times
\Big(
t_{f,i}^{{\rm imp}}(E) +
t_{f,i}^{N-{\rm exc}}(E) +
t_{f,i}^{\pi-{\rm exc}}(E) +
\{
{\rm exchange\ terms}
\}
\Big)  \sqrt{(2\omega)(2E_d(\bm{p}_d))}\ ,\nonumber \\
&&
\label{eq:amp_decomp}
\end{eqnarray}
where the sub-indices $f$ and $i$ stand for  
the final state $\bra{\pi(\bm{k},t_\pi)\, N_1(\bm{p}_1,s_1,t_1),N_2 (\bm{p}_2,s_2,t_2)}$,
and the initial state $\ket{\gamma(\bm{q},\lambda),\Psi_d(s_d)}$, respectively,
$\omega=|\bm{q}|$ is the photon energy, and 
for the laboratory frame,
\begin{eqnarray}
t_{f,i}^{{\rm imp}} &=&
\sqrt{2}
\sum_{s_1',t'_1}
\bra{\pi(\bm{k},t_\pi)\, N_1(\bm{p}_1,s_1,t_1)} 
t_{\pi N,\gamma N}(M_{\pi N_1})
\ket{\gamma(\bm{q},\lambda)\, N_1'(-\bm{p}_2 ,s_1',t'_1)}
\nonumber\\
&&\times
\inp{N_1'(-\bm{p}_2,s_1',t'_1)\, N_2 (\bm{p}_2,s_2,t_2)}{\Psi_d(s_d)}
\ ,
\label{eq:amp_imp}
\\
t_{f,i}^{N-{\rm exc}} &=&\sqrt{2}
\sum_{s_1',\tilde s_1',s_2',t'_1}
\int d\bm{l}\
\nonumber\\
&&\times
\bra{N_1(\bm{p}_1,s_1,t_1)\, N_2(\bm{p}_2,s_2,t_2)}
t_{NN,NN}(M_{N_1N_2}) 
\ket{\tilde N'_1(\bm{q}-\bm{k}+\bm{l},\tilde s'_1,t_1)\, N'_2(-\bm{l},s'_2,t_2)}
\nonumber\\
&&\times
{\bra{\pi(\bm{k},t_\pi)\, \tilde N_1'(\bm{q}-\bm{k}+\bm{l},\tilde s_1',t_1)} 
t_{\pi N,\gamma N}(W)
\ket{\gamma(\bm{q},\lambda)\, N_1'(\bm{l},s_1',t'_1)}
\over E-E_N(\bm{q}-\bm{k}+\bm{l})-E_N(-\bm{l})-E_\pi(\bm{k})+i\epsilon}
\nonumber\\
&&\times
\inp{N_1'(\bm{l},s_1',t'_1)\, N_2' (-\bm{l},s_2',t_2)}{\Psi_d(s_d)}
\ ,
\label{eq:amp_NN}
\\
t_{f,i}^{\pi-{\rm exc}} &=& \sqrt{2}
\sum_{s_1',s_2'}
\sum_{t_1',t_2',t'_\pi}
\int d\bm{l}
\bra{\pi(\bm{k},t_\pi)\, N_1(\bm{p}_1,s_1,t_1)}
t_{\pi N,\pi N}(M_{\pi N_1})
\ket{\pi(\bm{q}-\bm{p}_2+\bm{l},t'_\pi)\, N_1'(-\bm{l},s'_1,t_1')}
\nonumber\\
&&\times
{\bra{\pi(\bm{q}-\bm{p}_2+\bm{l},t'_\pi)\, N_2(\bm{p}_2,s_2,t_2)} 
t_{\pi N,\gamma N}(W)
\ket{\gamma(\bm{q},\lambda)\, N_2'(\bm{l},s_2',t_2')}
\over 
E-E_N(\bm{p}_2)-E_N(-\bm{l})-E_{\pi}(\bm{q}-\bm{p}_2+\bm{l})+i\epsilon}
\nonumber\\
&&\times
\inp{N_1'(-\bm{l},s_1',t_1')\, N_2' (\bm{l},s_2',t_2')}{\Psi_d(s_d)}
\ .
\label{eq:amp_MN}
\end{eqnarray}
The exchange terms in Eq.~(\ref{eq:amp_decomp}) can be obtained 
from Eqs.~(\ref{eq:amp_imp})-(\ref{eq:amp_MN}) 
by flipping the overall sign 
and interchanging 
all subscripts 1 and 2 for the nucleons in the intermediate and final $\pi NN$ states.
Here, the deuteron state  with spin projection $s_d$ is denoted as $\ket{\Psi_d(s_d)}$;
$\ket{N(\bm{p},s,t)}$ the nucleon state with momentum $\bm{p}$ and spin and isospin projections 
$s$ and $t$;
$\ket{\gamma(\bm{q},\lambda)}$ the photon state with momentum $\bm{q}$ and polarization $\lambda$;
$\ket{\pi(\bm{k},t_\pi)}$ the pion state with momentum $\bm{k}$ and 
the isospin projection $t_\pi$.
The total energy in the laboratory frame, $E$, is given by
$E=\omega+m_d$ where $m_d$ is the deuteron mass.
The invariant masses for the matrix elements of the
two-body subprocesses are calculated according to the momentum variables specified in
Fig.~\ref{fig:diag} and 
 $E-E_{\rm spectator}$ of the spectator approximation defined by Eq.~(\ref{eq:spect}):
\begin{eqnarray}
M_{\pi N_1}&=&\sqrt{[E_\pi(\bm{k})+E_{N}(\bm{p}_1)]^2-(\bm{k}+\bm{p}_1)^2}\ ,\\
M_{N_1 N_2}&=&\sqrt{[E_{N}(\bm{p}_1)+E_{N}(\bm{p}_2)]^2-(\bm{p}_1+\bm{p}_2)^2}\ ,\\
\label{eq:W}
W&=&\sqrt{[E-E_{N}(-\bm{l})]^2-(\bm{l}+\bm{q})^2}\ .
\end{eqnarray}
To be consistent with the chosen CD-Bonn $NN$ potential~\cite{cdbonn}
for generating the $NN$ scattering amplitudes and the deuteron bound state,
the  two-nucleon energy in the propagator of the nucleon  re-scattering amplitude 
$t_{NN,NN}$ in Eq.~(\ref{eq:amp_NN}),
is calculated with  a non-relativistic approximation
\begin{eqnarray}
E_N(\bm{q}-\bm{k}+\bm{l})+E_N(-\bm{l}) \sim 
\sqrt{(2m_N)^2 + (\bm{q}-\bm{k})^2} + {(\bm{q}/2-\bm{k}/2+\bm{l})^2\over
m_N} \ .
\end{eqnarray}

The amplitudes $\bra{\pi N} t_{\pi N,\gamma N} \ket{\gamma N'}$ of the 
pion photoproduction and $\bra{\pi N}t_{\pi N,\pi N}\ket{\pi'N'}$ of the
pion-nucleon scattering 
in Eqs.~(\ref{eq:amp_imp})-(\ref{eq:amp_MN}) are first generated from
the ANL-Osaka model by 
solving the coupled-channel equation of Eq.~(\ref{eq:cceq}) in the two-body
CM frame. The resulting matrix elements are then boosted to the considered
$\gamma$-deuteron frame. 
Here we follow the approach of Ref.~\cite{polyzou} based on the instant form of 
relativistic quantum mechanics~\cite{kei-pol}.
The same frame-transformation  procedure is also needed to calculate
the matrix element $\bra{NN}t_{NN,NN}\ket{NN}$ of $NN$ scattering in Eq.~(\ref{eq:amp_NN}). 
The  formulas for calculating these matrix elements in 
the $\gamma$- deuteron frame are given in Appendix~\ref{app1}.

Here we note an important difference between our dynamical model approach and the approach
of Refs.~\cite{tara,tara-1}. In the loop-integrations of Eqs.~(\ref{eq:amp_NN}) and
(\ref{eq:amp_MN}), the two-body matrix
elements can be off-energy-shell and are calculated exactly from the ANL-Osaka
model and the CD-Bonn potential as explained above. 
The equations for calculating FSI in Refs.~\cite{tara,tara-1} are similar to our
expressions, but the two-body matrix elements are taken as
their on-shell values and are
taken out of the loop-integrations.
For $NN\rightarrow NN$, they include a monopole form factor of Ref.~\cite{lev06}
to account for the off-shell effect. 
These simplifications greatly reduce  the computation task.
We will examine the on-shell approximation within our formulation in Sec.~\ref{sec:result}.

\subsection{Cross section formula}
\label{sec:cross_section}

The reaction cross section 
for $\gamma(\bm{q})+ d(\bm{p}_d)\to \pi(\bm{k})+ N_1(\bm{p}_1)+N_2(\bm{p}_2)$ is defined by
\begin{eqnarray}
d\sigma&=&\frac{(2\pi)^4}{|\bm{v}_{rel}|}
\delta^{(3)}(\bm{p}_d+\bm{q}-\bm{p}_1-\bm{p}_2-\bm{k})
\delta(E_d(\bm{p}_d)+\omega -E_N(\bm{p}_1)-E_N(\bm{p}_2)-E_\pi(\bm{k}))
\nonumber \\
&&\times d\bm{p}_1d\bm{p}_2d\bm{k} \left[\frac{m_N}{E_N(\bm{p}_1)}
\frac{m_N}{E_N(\bm{p}_2)}\frac{1}{2E_\pi(\bm{k})}\right]
\,|M_{f,i}(E)|^2\, \left[\frac{1}{2E_d(\bm{p}_d)} \frac{1}{2\omega}\right] \ ,
\label{eq:dsigma-g}
\end{eqnarray}
where 
$\bm{v}_{rel}$ is the relative velocity of the initial $\gamma$-$d$ system,
and the Lorentz invariant amplitude $M_{f,i}(E)$ has been defined by 
Eqs.~(\ref{eq:amp_decomp})-(\ref{eq:amp_MN}).
The unpolarized differential cross section 
with respect to the pion emission angle in the
laboratory frame ($\bm{p}_d=0$) derived from Eq.~({\ref{eq:dsigma-g}) can be written as
\begin{eqnarray}
{d\sigma \over d\Omega} 
\equiv {d\sigma \over d\Omega_{\bm{k}}}
=\frac{1}{6}\sum_{\lambda,s_d}\frac{d\sigma(\lambda,s_d)}{d\Omega_{\bm{k}}}\ ,
\label{eq:dsigp}
\end{eqnarray}
where $\lambda$ is the photon polarization, $s_d$ is the $z$-component of the deuteron spin, and
\begin{eqnarray}
\frac{d\sigma(\lambda,s_d)}{d\Omega_{\bm{k}}}=
\int dM_{NN}\frac{d\sigma(\lambda,s_d)}{d\Omega_{\bm{k}}dM_{NN}} \ ,
\label{eq:dsigp-1}
\end{eqnarray}
with 
\begin{eqnarray}
\frac{d\sigma(\lambda,s_d)}{d\Omega_{\bm{k}}dM_{NN}}
= 
\sum_{s_1,s_2}
\frac{(2\pi)^4}{4\omega}\frac{1}{2E_d(\bm{p}_d)}
\int d\Omega_{\bm{p}_{NN}} {p_{NN} k^2 m_N^2\over |k E -
\bm{q}\!\cdot\!\hat{k}\, E_\pi(\bm{k})|}\;
|M_{f,i}(E)|^2 \ .
\label{eq:dsigp-nn}
\end{eqnarray}
Here $E=m_d+\omega$ is the total energy in the laboratory frame,  
$\bm{p}_{NN}$ the momentum of $N_1$ in the 
$N_1N_2$ CM frame, and 
$M_{NN}$ the invariant mass of the two  nucleons in the final $\pi NN$ state.
The magnitudes of the momenta are simply denoted, for example, by 
$k\equiv |\bm{k}|$, and $\hat{k}\equiv \bm{k}/|\bm{k}|$. 
To evaluate the Lorentz invariant  amplitude $M_{f,i}(E)$ using 
Eqs.~(\ref{eq:amp_decomp})-(\ref{eq:amp_MN}),
we need to express each outgoing momentum in Fig.~\ref{fig:diag}
in terms of $\bm{k}$, $\hat{p}_{NN}$ and
$M_{NN}$. This can be done by using the standard Lorentz transformation.
For a given $\bm{k}$, $M_{NN}$, and the angles $\Omega_{\bm{p}_{NN}}$, 
we can calculate them by
\begin{eqnarray}
p_{NN}&=&|\bm{p}_{NN}|=\sqrt{{1\over 4}M^2_{NN}-m^2_N} \label{eq:lt-1}\ , \\
\bm{p}_1&=&\bm{p}_{NN}+\gamma\left(\frac{\gamma}{1+\gamma} 
\bm{\beta}\cdot\bm{p}_{NN}-E_N(\bm{p}_{NN})\right)\bm{\beta}\ ,  \\
\bm{p}_2&=&\bm{P}_{NN}-\bm{p}_1\ , 
\end{eqnarray}
with
\begin{eqnarray}
\bm{P}_{NN}&=&\bm{q}-\bm{k}\ , \\
\bm{\beta}&=&- \frac{\bm{P}_{NN}}{[M^2_{NN}+\bm{P}_{NN}^2]^{1/2}}\ , \\
 \gamma&=&\frac{1}{\sqrt{1-\bm{\beta}^2}} \ .
\label{eq:lt-2}
\end{eqnarray}
We will also present results of 
the differential cross section in the 
the CM frame of the $\gamma$-$d$ system.
We find that 
\begin{eqnarray}
{d\sigma \over d\Omega_c}
\equiv {d\sigma \over d\Omega_{\bm{k}_c}}
=\frac{1}{6}\sum_{\lambda,s_d}\frac{d\sigma(\lambda,s_d)}{d\Omega_{\bm{k}_c}}\ ,
\label{eq:dsigp_CM}
\end{eqnarray}
where
\begin{eqnarray}
\frac{d\sigma(\lambda,s_d)}{d\Omega_{\bm{k}_c}}=
\int dM_{NN}\frac{d\sigma(\lambda,s_d)}{d\Omega_{\bm{k}_c}dM_{NN}} \ ,
\label{eq:dsigp_CM-1}
\end{eqnarray}
with
\begin{eqnarray}
\frac{d\sigma(\lambda,s_d)}{d\Omega_{\bm{k}_c}dM_{NN}}=
\sum_{s_1,s_2}{1\over |\bm{v}_{{\rm rel}, c}|}
\frac{(2\pi)^4}{4\omega_c}\frac{1}{2E_d(\bm{q}_c)}
\int  d\Omega_{\bm{p}_{NN}} {p_{NN} k_c m_N^2\over 
E_c
}\;
|M_{f,i}(E_c)|^2 \ ,
\label{eq:dsigp_CM-nn}
\end{eqnarray}
where $(\omega_c,\bm{q}_c)$ and $E_c$ can be obtained from 
$(\omega,\bm{q})$ and $E$ in the laboratory frame by the Lorentz transformation with
$\bm{\beta}=\bm{q}/{(\omega+m_d)}$.
The momentum variables for the calculations of the invariant amplitudes using 
Eqs.~(\ref{eq:amp_decomp})-(\ref{eq:amp_MN}) can be obtained
from the same Eqs.~(\ref{eq:lt-1})-(\ref{eq:lt-2}), but setting 
$\bm{P}_{NN}=-\bm{k_c}$.

We will also present results for three polarization observables of current 
interest.
We choose 
the $z$-axis along the incident photon direction, 
$\bm{q}=(q_x,q_y,q_z)=(0,0,q)$, 
and the $y$-axis along the vector product, $\hat q \times \hat k$.
Then, the photon asymmetry $\Sigma$ for the $d(\gamma,\pi)NN$ reaction is defined by
\begin{eqnarray}
\Sigma_{\gamma d} =\frac{d\sigma(\lambda = 2)-d\sigma(\lambda = 1)}
{d\sigma(\lambda = 2)+d\sigma(\lambda = 1)} \ ,
\label{eq:sigma}
\end{eqnarray}
where 
$d\sigma(\lambda)$ denotes 
either of differential cross sections of 
Eqs.~(\ref{eq:dsigp}) or (\ref{eq:dsigp_CM}),
but the initial photon polarization is fixed to $\lambda$ instead of the average.
The components $\lambda=1,2$ for the linear photon polarization
in Eq.~(\ref{eq:sigma}) are
\begin{eqnarray}
\bm{\epsilon}_{\lambda=1}=(1,0,0) \ , \quad
 \bm{\epsilon}_{\lambda=2}=(0,1,0) \ .
\end{eqnarray}
The asymmetry $E$ of the circularly polarized photons on 
a deuteron target polarized along the $z$-axis ($s_d=+1$) is defined by
\begin{eqnarray}
E_{\gamma d}=\frac{d\sigma(\lambda= 3,s_d=+1)-d\sigma(\lambda = 4,s_d=+1)}
{d\sigma(\lambda = 3,s_d=+1)+d\sigma(\lambda = 4,s_d=+1)} \ ,
\label{eq:E}
\end{eqnarray}
where 
$d\sigma(\lambda,s_d=+1)$ denotes 
either of differential cross sections of 
Eqs.~(\ref{eq:dsigp-1}) or (\ref{eq:dsigp_CM-1}); 
the deuteron spin orientation is $s_d=+1$ and the components
$\lambda=3,4$ for the circular photon polarization are defined as
\begin{eqnarray}
\bm{\epsilon}_{\lambda=3}=\frac{+1}{\sqrt{2}}(1,-i,0) \ , \quad
\bm{\epsilon}_{\lambda=4}=\frac{-1}{\sqrt{2}}(1,i,0) \ .
\label{eq:pol_E}
\end{eqnarray}
The asymmetry $G$ of the linearly polarized photons on a 
polarized deuteron is defined by
\begin{eqnarray}
G_{\gamma d} =\frac{d\sigma(\lambda = 5,s_d=+1)-d\sigma(\lambda = 6,s_d=+1)}
{d\sigma(\lambda = 5,s_d=+1)+d\sigma(\lambda = 6,s_d=+1)} \ ,
\label{eq:G}
\end{eqnarray}
where the components $\lambda=5,6$ for the photon polarization  are                              
\begin{eqnarray}
 \bm{\epsilon}_{\lambda=5}=\frac{1}{\sqrt{2}}(1,1,0) \ , \quad
\bm{\epsilon}_{\lambda=6}=\frac{1}{\sqrt{2}}(1,-1,0) \ .
\end{eqnarray}

\section{Results} 
\label{sec:result}

In our calculations of the amplitudes Eqs.~(\ref{eq:amp_imp})-(\ref{eq:amp_MN}), 
the  $\gamma N, \pi N \rightarrow \pi N$ partial-wave amplitudes  up to
and including $L=5$ 
($L$: the $\pi N$ orbital angular momentum) 
generated from the ANL-Osaka model of Refs.~\cite{knls13,knls16}
are taken into account.
To calculate the loop-integrations 
in Eqs.~(\ref{eq:amp_NN}) and (\ref{eq:amp_MN}) 
for the $N$- and $\pi$-exchange mechanisms, respectively, 
the off-energy-shell two-body matrix elements are 
generated from solving the scattering equation Eq.~(\ref{eq:cceq}) for
the meson-nucleon scattering
and a similar Lippmann-Schwinger equation of Eq.~(\ref{eq:LS}) for the $NN$ scattering.
The $NN\rightarrow NN$ partial wave amplitudes up to and including
$J=3$ ($J$: the $NN$ total angular momentum) 
generated from
the CD-Bonn potential are included in the calculations.

We first compare our predictions with the available data
of $d(\gamma,\pi^0)pn$ and $d(\gamma,\pi^-)pp$ reactions. We then examine whether
our results can be improved if the employed $\gamma N \rightarrow \pi N$ model
of Ref.~\cite{knls16} is adjusted to also fit the recent Mainz
data~\cite{mami1,mami2} of $\gamma n\rightarrow \pi^0n$.
 We also examine the roles of the off-energy-shell and FSI effects
 in explaining the data.
The differences with the approach of Refs.~\cite{tara,tara-1}
will also be  discussed.

\begin{figure}[h]
\begin{center}
\includegraphics[clip,width=1.\textwidth]{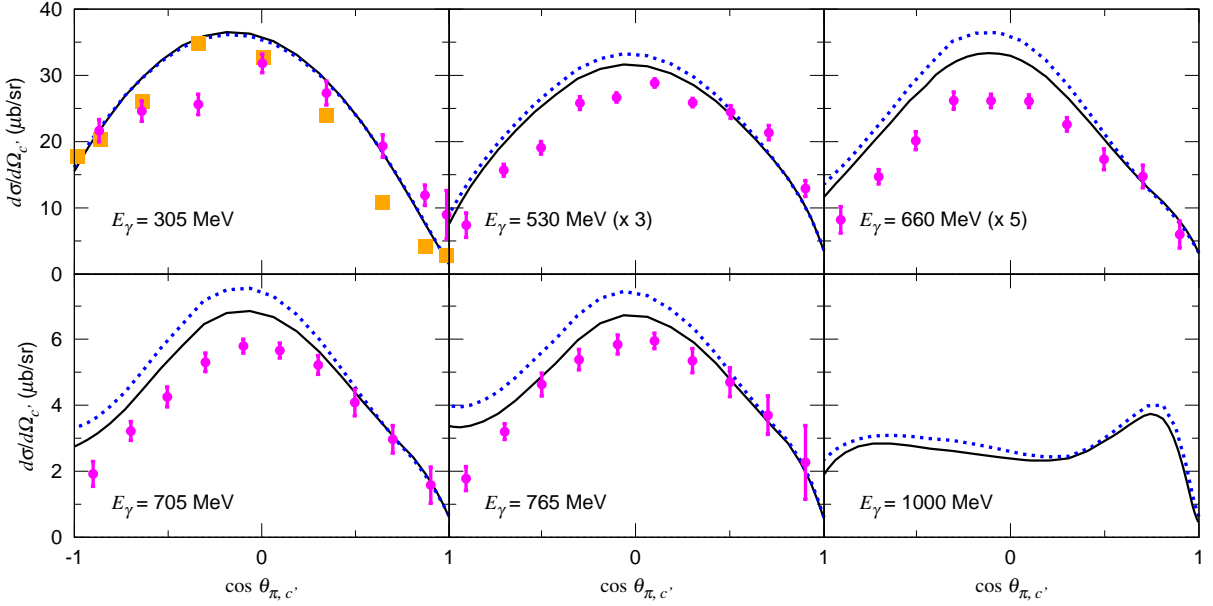}
\caption{
Differential cross sections for $d(\gamma,\pi^0)pn$
calculated with  the $\gamma N \rightarrow \pi N$  amplitudes from 
 the AO model~\cite{knls13,knls16} (blue dotted curves) and
 the new fits including the recent Mainz data~\cite{mami1,mami2} of $\gamma n \rightarrow \pi^0 n$  (black solid curves).  
The cross sections are scaled by the factor in the parenthesis when it
 is given.
The pion emission angle $\theta_{\pi,c'}$ 
is defined in the frame where the photon momentum
$\bm{q}_{c'}$ and the deuteron momentum 
$\bm{p}_{d,c'}$ are related by $\bm{p}_{d,c'}=-2\bm{q}_{c'}$. 
The photon energies indicated in the panels ($E_\gamma$) are those in the
 laboratory frame.
Data are from Ref.~\cite{gd-pi0pn_data} (magenta circles)
and Ref.~\cite{gd-pi0pn_data2} (orange squares).
}
 \label{fig:gd-pi0pn-dat}
\end{center}
\end{figure}
\begin{figure}[h]
\begin{center}
\includegraphics[clip,width=1.\textwidth]{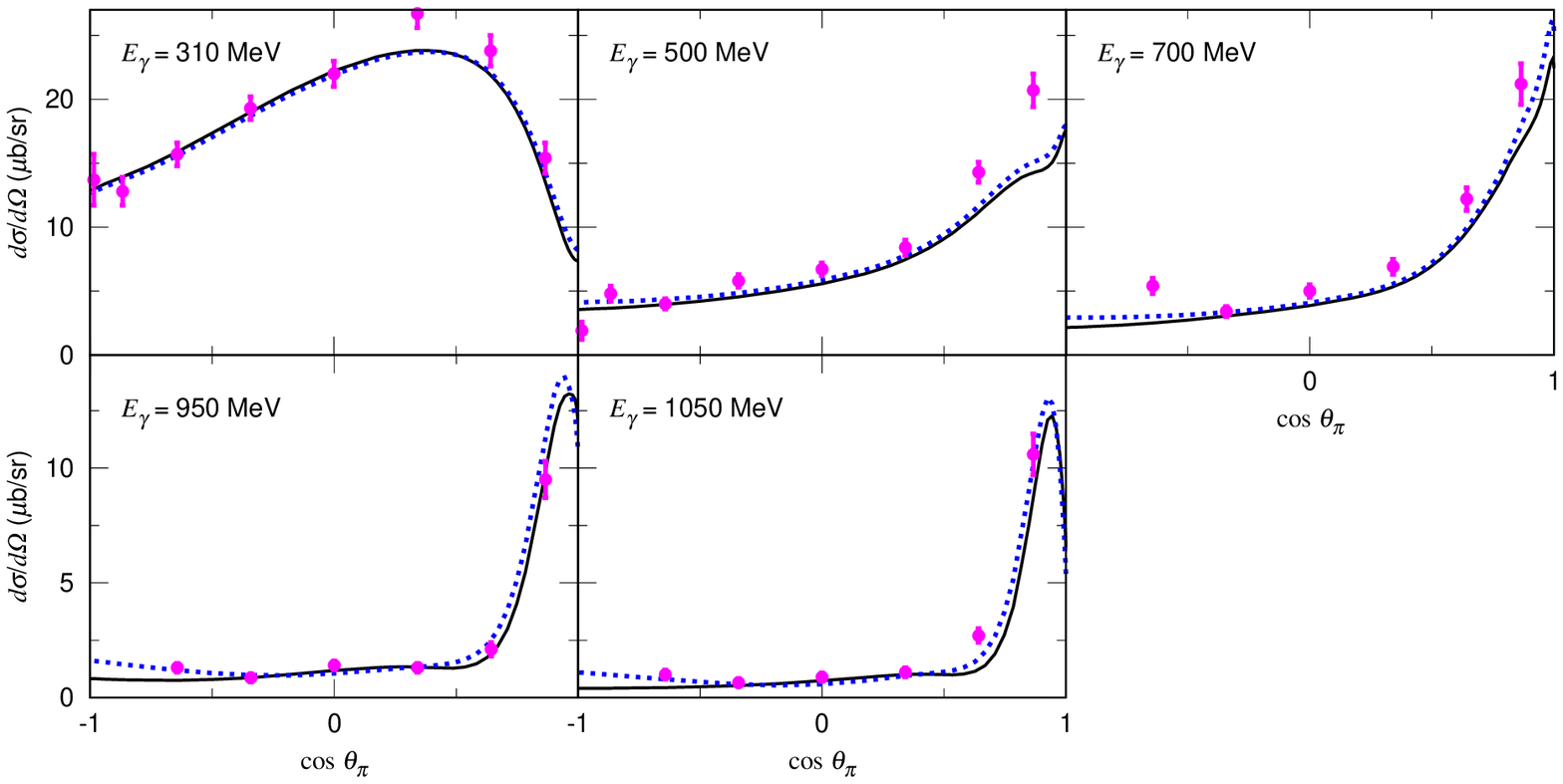}
\caption{
Differential cross sections for $d(\gamma,\pi^-)pp$ in the
 laboratory frame. 
Data are from Ref.~\cite{gd-pimpp_data}.
The other features are the same as those in Fig.~\ref{fig:gd-pi0pn-dat}.
}
 \label{fig:gd-pimpp-dat}
\end{center}
\end{figure}
\subsection{Comparisons with data of $d(\gamma,\pi)NN$ reactions}

We have found that our predictions agree reasonably well with the available data of
$d(\gamma,\pi^0)pn$ and $d(\gamma,\pi^-)pp$ reactions.
In Figs.~\ref{fig:gd-pi0pn-dat} and \ref{fig:gd-pimpp-dat}, 
we show some typical comparisons of the data for unpolarized differential cross sections
and our results (blue dotted curves) 
using the amplitudes generated from
the ANL-Osaka model.
Clearly there are some discrepancies with the data, in particular for $d(\gamma,\pi^0)np$
in the $E_\gamma=$~500--750~MeV region
($E_\gamma$ is the photon energy in the laboratory frame).
One possible source of the discrepancies is that,
when determining the parameters for the ANL-Osaka model, 
we used rather scarce $\gamma n \rightarrow \pi^0n$ dataset that are 
extracted from $d(\gamma,\pi^0)np$ data.
To see whether this is the case, we have extended the fit of Ref.~\cite{knls16}
to also include recently extracted data for 
 $ \gamma n\rightarrow\pi^0 n$ from Mainz~\cite{mami1,mami2}, and also
recent data for
$\gamma n \rightarrow \pi^-p$ from JLab~\cite{clas1,clas2}. However,
the amount of these new data are much less than the total number of
the world data for $\pi N, \gamma N\rightarrow \pi N, \eta N, K\Lambda,K\Sigma$
included in the fit~\cite{knls13,knls16} for the ANL-Osaka model.
We thus do not expect that
the resonance parameters and the hadronic parameters 
determined in Ref.~\cite{knls16} will be changed significantly
by including the recent $\gamma n\to\pi N$ data, and therefore vary
only the parameters for the bare $\gamma n \rightarrow N^*$ couplings in
the new fit.

\begin{figure}[h]
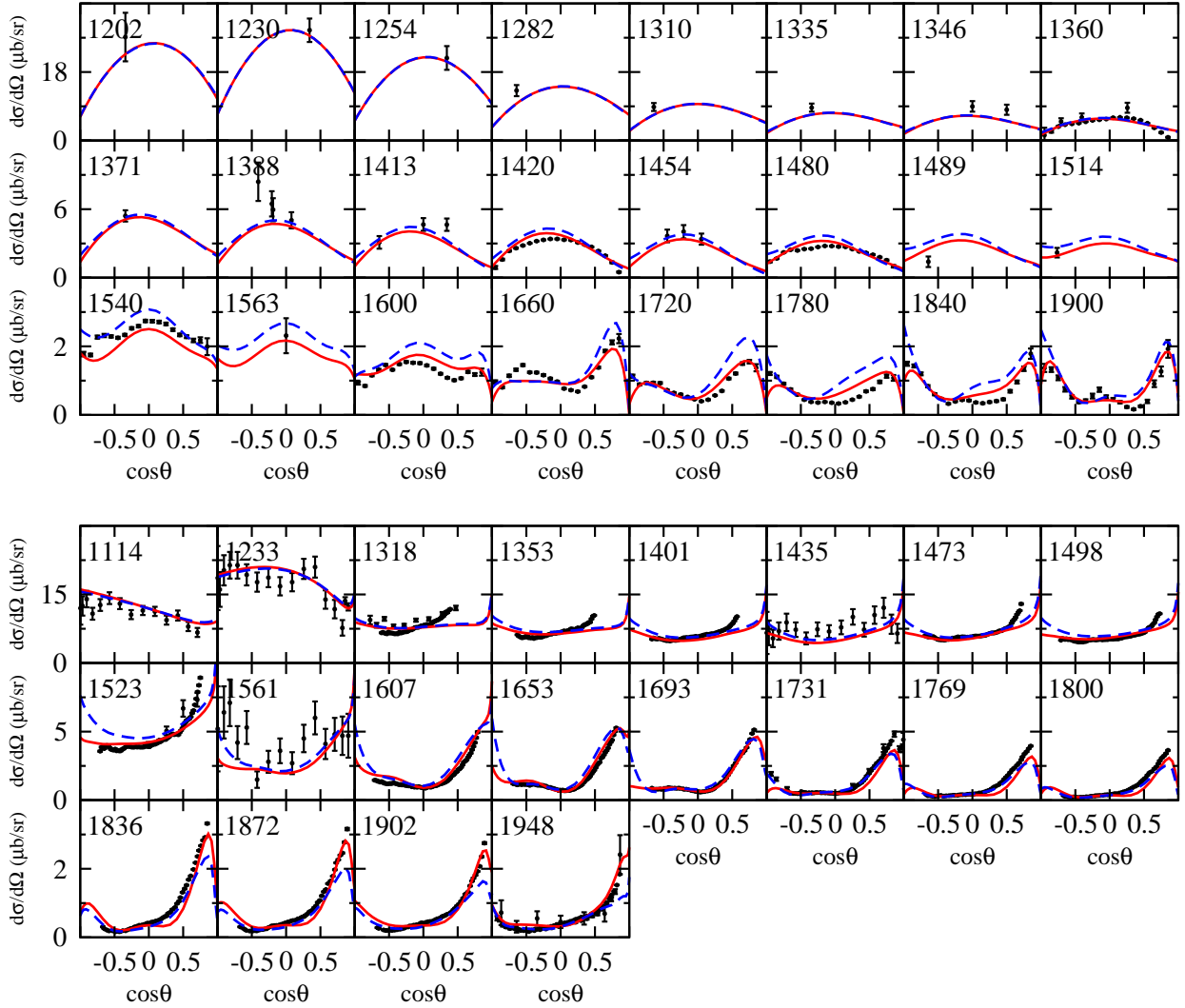

\begin{center}
\includegraphics[clip,width=1.\textwidth]{gnpi0n-dc-reduced}

\vspace*{5mm}
\includegraphics[clip,width=1.\textwidth]{gnpimp-dc-reduced}
\caption{
Differential cross sections for $\gamma n \rightarrow \pi^0 n$ (upper panel)
and for  $\gamma n \rightarrow \pi^- p$ (lower panel) in the CM frame. 
The number in each of the panels is the total energy in unit of MeV.
The red solid curves are from the new fit while the blue dashed curves
 are from the AO model~\cite{knls16}.
The data are from Ref.~\cite{mami1} for $\gamma n \rightarrow\pi^0 n$
and Ref.~\cite{clas2} for $\gamma n \rightarrow \pi^- p$
in addition to those from 
the database of the INS DAC Services~\cite{INS_DAC}.
}
 \label{fig:mami1}
\end{center}
\end{figure}
To be consistent with the procedure in constructing the ANL-Osaka model,
the calculations for the fits include $L=0-10$ partial waves.
In addition to the data 
of $\pi p, \gamma p \rightarrow \pi N, \eta N, K\Lambda, K\Sigma$ and
$\gamma n \rightarrow \pi^0 n,\pi^-p$  included in the fits in
Ref.~\cite{knls16}, the recent data~\cite{mami1,mami2,clas1,clas2} of
the unpolarized differential cross sections
and $E$ for $\gamma n \rightarrow \pi^0 n, \pi^-p$ are included in the new fits.
In Fig.~\ref{fig:mami1}, we show the comparisons of our
results with the data of the differential cross sections.
The fits to the  data which  are not shown in Fig.~\ref{fig:mami1}
are of similar quality and hence are omitted to simplify the presentation.
Clearly,  the results from this new  fit (red solid curves) 
and those from the ANL-Osaka model (blue dashed curves) are very similar.
On the other hand, 
we see in Fig.~\ref{fig:clas1} 
some large differences between the fits and the data for
the polarization observable $E$. 
 Further improvements are possible only
if we also vary
the non-resonant parameters around the values determined
in Ref.~\cite{knls16}. This will be worthwhile to pursue when
the data for other spin observables, such as $G$ and $\Sigma$,
become available. 
For our present purposes, the quality of the fit shown in 
Figs.~\ref{fig:mami1}-\ref{fig:clas1}
is sufficiently good.
We note here that
the higher partial waves ($L=6-10$),
which include only non-resonant Born amplitudes,
just slightly change the shape of the angular
distributions for $\gamma n\to\pi^- p$ in the forward pion kinematics, 
and has little effects on  $\gamma n\to\pi^0 n$ .
These weak higher partial wave amplitudes are therefore neglected in all of our calculations
for the $d(\gamma,\pi)NN$ reactions.
\begin{figure}[h]
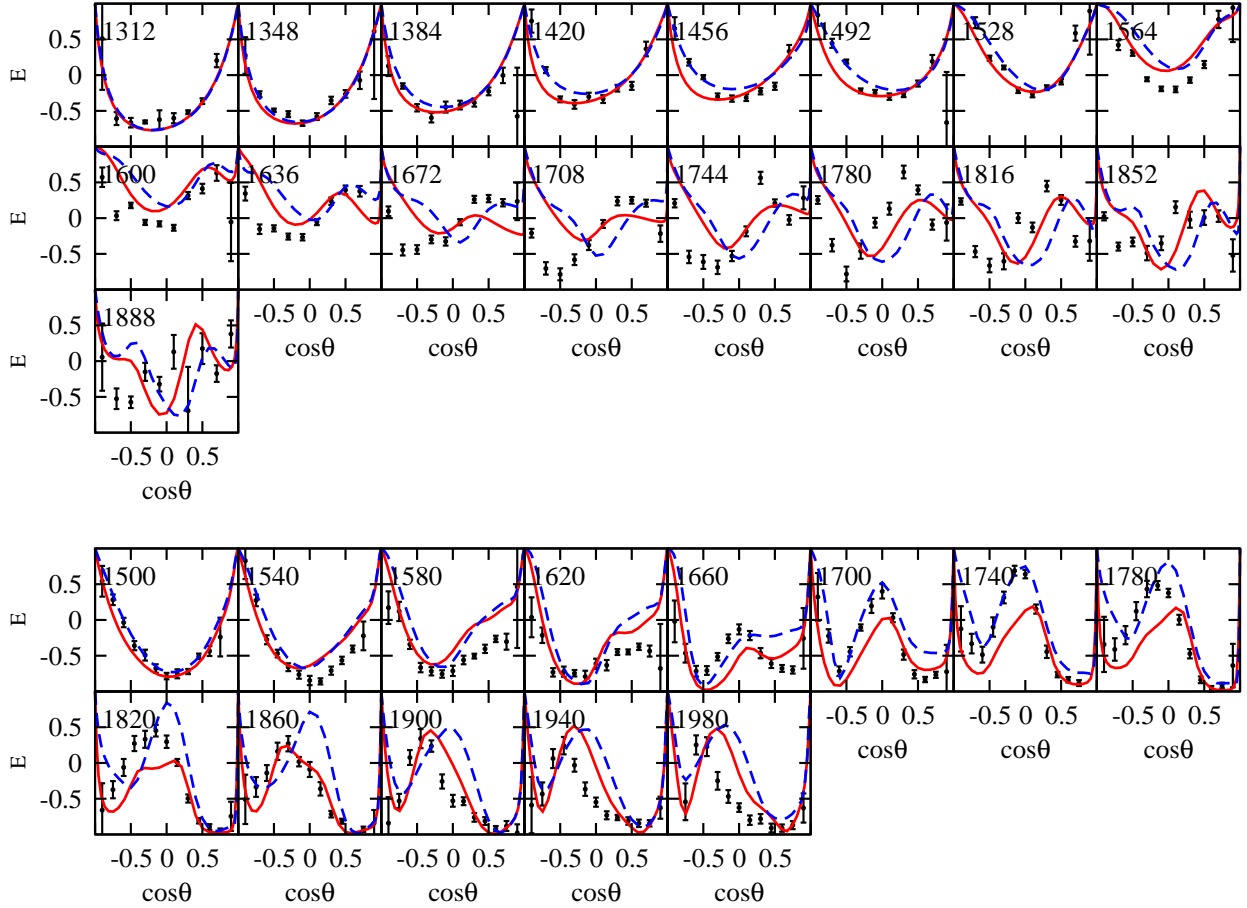

\begin{center}
\includegraphics[clip,width=1.0\textwidth]{gnpi0n-e}

\vspace*{5mm}
\includegraphics[clip,width=1.0\textwidth]{gnpimp-e}
\caption{
The polarization observable $E$  for $\gamma n \rightarrow \pi^0n$
(upper panel) and 
 $\gamma n \rightarrow \pi^-p$ (lower panel)
The data are from Ref.~\cite{mami2} for $\gamma n \rightarrow\pi^0 n$
and Ref.~\cite{clas1} for $\gamma n \rightarrow \pi^- p$.
The other features are the same as those in Fig.~\ref{fig:mami1}.}
 \label{fig:clas1}
\end{center}
\end{figure}

With the parameters from the new fits, the calculated differential cross sections for
$d(\gamma, \pi^0)pn$ and $d(\gamma, \pi^-)pp$ are the solid curves in 
Figs.~\ref{fig:gd-pi0pn-dat} and \ref{fig:gd-pimpp-dat}. Comparing with the
 blue dotted curves from using the ANL-Osaka amplitudes, we see that the agreements with
the data for $d(\gamma,\pi^0)pn $ in the region $E_\gamma = 500-700$ MeV are
improved, but some significant discrepancies with the data remain. 
It could be due to the neglect of higher order terms
in our multiple scattering calculations such as $\rho$-, $\sigma$-,
 and $\Delta$-exchange terms. On the other hand, we must examine critically
the procedures used in extracting the $\gamma n \rightarrow \pi^0 n, \pi^-p$
cross sections from the data of $d(\gamma,\pi)NN$.
The results presented in the rest of this paper are from using the parameters from
the new fit.

\subsection{Off-shell  effects}

\begin{figure}[h]
\begin{center}
\includegraphics[clip,width=.69\textwidth]{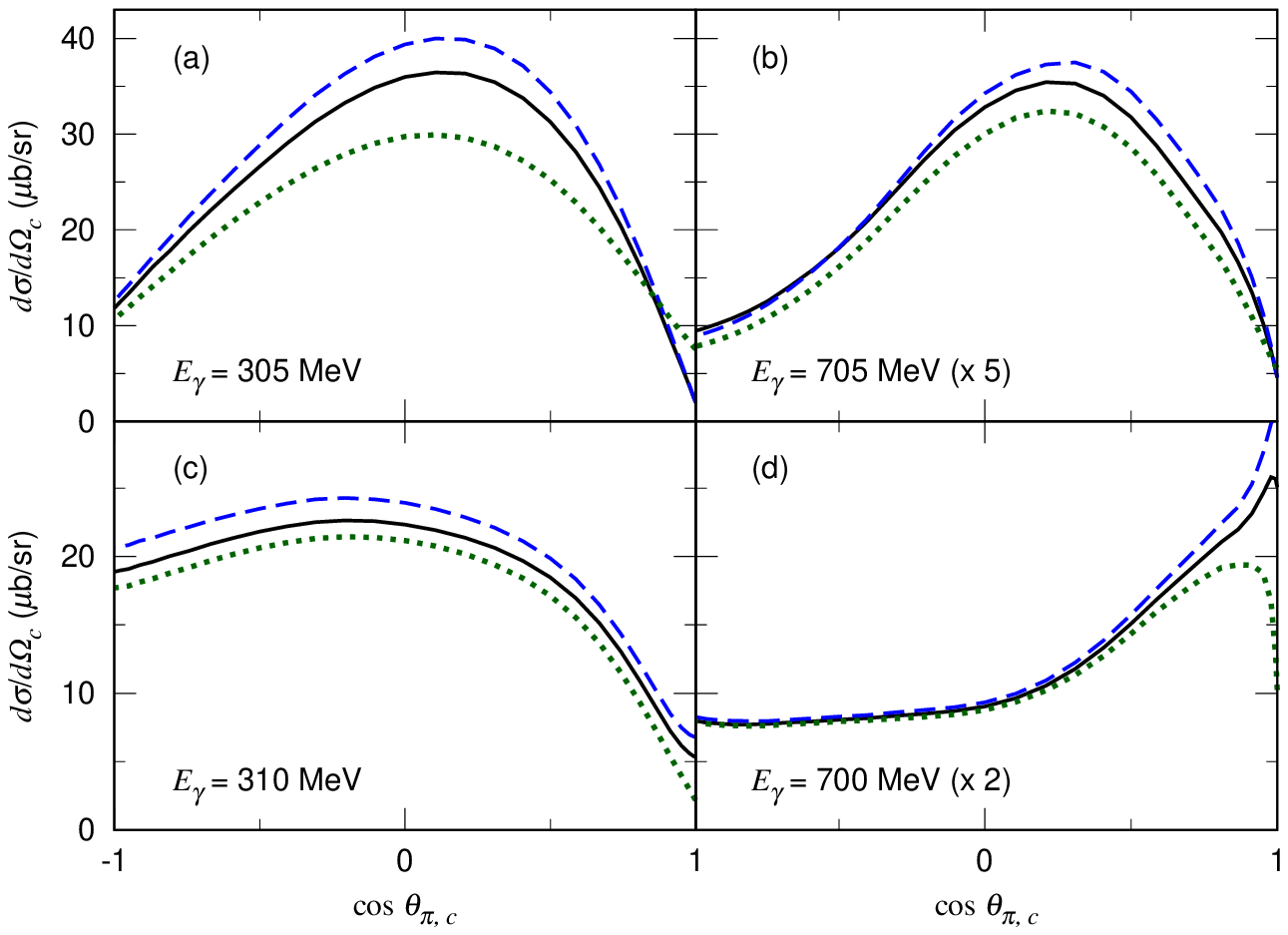}
\caption{The off-shell effects on differential cross sections for
$d(\gamma,\pi^0)pn$ (a,b) and
$d(\gamma,\pi^-)pp$ (c,d) in the CM frame.
The photon energy and the scaling factor
are indicated in each figure.
Solid curves: full calculations;
 dashed curves: the $\gamma N\to\pi N$, $NN\to NN$, and $\pi N\to\pi N$ amplitudes
are put on-shell;
dotted curves:  only pole terms are kept in loop integrations
 [see Eq.~(\ref{eq:pole})].
}
 \label{fig:gd-piNN-offshell}
\end{center}
\end{figure}
In Sec.~\ref{sec:model}, we show that the two-body matrix elements in the calculations
of $d(\gamma,\pi)NN$ amplitudes can be off-energy-shell within the Hamiltonian
formulation where all particles are on their mass-shell.
In Fig.~\ref{fig:gd-piNN-offshell}, we see that, if we set all two-body matrix elements
in Eqs.~(\ref{eq:amp_imp})--(\ref{eq:amp_MN}) 
to their on-shell values, the full results (solid curves) are changed to
the blue dashed curves. 
In general, the effects are more significant at low energies 
and less at higher energies, as illustrated at energies near 300 and 700 MeV in
Fig.~\ref{fig:gd-piNN-offshell}. The differences between the solid and blue dashed curves
suggest that the $\gamma n \rightarrow \pi^-p$ extracted
using the on-shell calculation results of Refs.~\cite{tara} are reasonable,
but could be modified if the off-shell
effects are included in their formulation. The off-shell effects are however model dependent and
one must use well-established physics to determine them. In our approach, the off-shell effects
are determined by the meson-exchange mechanisms in the ANL-Osaka model and
CD-Bonn $NN$ potential. Similar considerations must also be taken in other approaches. 

If we further keep only the on-shell pole terms of the propagators
by setting 
\begin{eqnarray}
\frac{1}{E-E_N(\bm{p}_1)-E_N(\bm{p}_2)-E_\pi(\bm{k})+i\epsilon}\rightarrow 
-i\pi \delta(E-E_N(\bm{p}_1)-E_N(\bm{p}_2)-E_\pi(\bm{k})) \ ,
\label{eq:pole}
\end{eqnarray}
in the loop integrations
in Eqs.~(\ref{eq:amp_NN}) and (\ref{eq:amp_MN}),
we obtain the green dotted curves. Clearly, they are significantly different from
the solid curves of the full calculations including the off-shell
effects, in particular,
at $E_\gamma=$ 305 MeV for $d(\gamma,\pi^0)pn$, and in the forward angles 
($\cos\theta_{\pi, c}\rightarrow 1$)
at $E_\gamma=$ 700 MeV for $d(\gamma,\pi^-)pp$ .
Here we mention that Eq.~(\ref{eq:pole}) is sometimes used 
to reduce the computation effort for the loop-integration.
The differences between the solid and dotted curves in 
Fig.~\ref{fig:gd-piNN-offshell} give some
estimates on the accuracy of such a
simplified approach which neglects the off-shell effects
completely.

\subsection{Final state interaction effects}

In all of the previous investigations of $d(\gamma,\pi^0)pn$ reaction, the $N$-exchange  
term illustrated in Fig.~\ref{fig:diag}(b)
is found to dominant the FSI.
This can be examined by using Eq.~(\ref{eq:dsigp_CM-nn}) to calculate
the dependence of the differential cross sections on the $NN$ invariant mass $M_{NN}$.
In the upper half of Fig.~\ref{fig:gd-pi0pn-NN}, we see 
at a small angle of $10^\circ$ that
the main contributions to the differential cross sections of $d(\gamma,\pi^0)pn$
at $E_\gamma=305$~MeV are in the region close to the $NN$ threshold
$(M_{NN}-2\, m_N\sim 0)$ where the $NN$ scattering is dominated by the
$^3S_1+^3D_1$ partial wave which is orthogonal to the deuteron bound state. 
Consequently,
the $NN$ FSI greatly reduces the cross section of the impulse 
($t^{\rm imp}$) term (dashed curve) to
that of the $t^{\rm imp}+t^{N-{\rm exc}}$ terms (dotted curve).
When the pion-exchange term ($t^{\pi-{\rm exc}}$) is included, we obtain the solid curve which
is almost indistinguishable from the dotted curve.
As the scattering angle increases to $60^\circ$, 
the peak position of the $M_{NN}$ spectrum is
shifted to the larger invariant mass region where the reduction due to the $NN$ FSI
is still large, but much less than that in the forward angles. The FSI effects become
weak at large angles $100^\circ$ and $150^\circ$ since the large portion
of the $NN$ spectrum is
in the region far away from the $NN$ threshold.
\begin{figure}[t]
\begin{center}
\includegraphics[clip,width=1\textwidth]{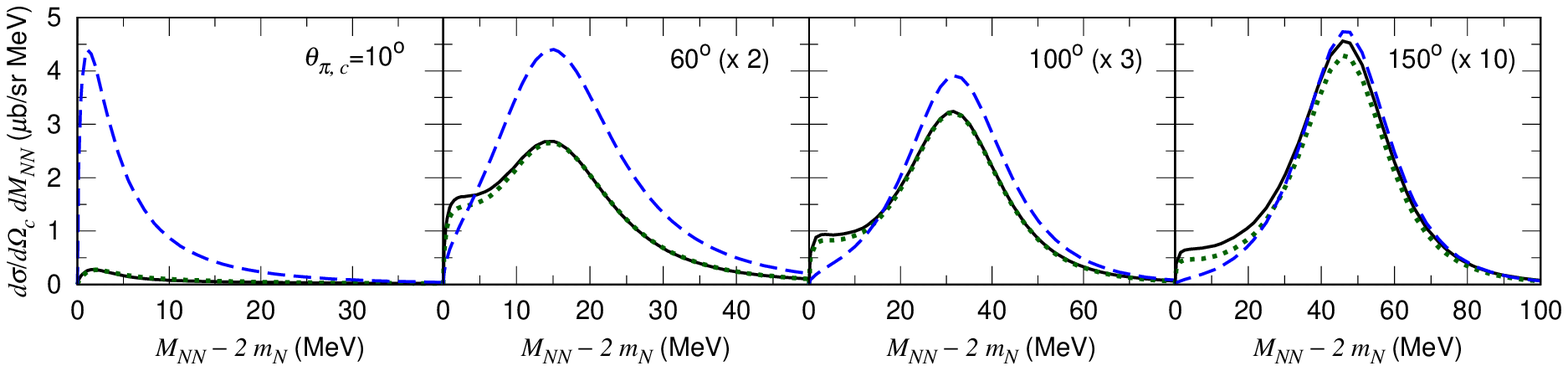}
\includegraphics[clip,width=1\textwidth]{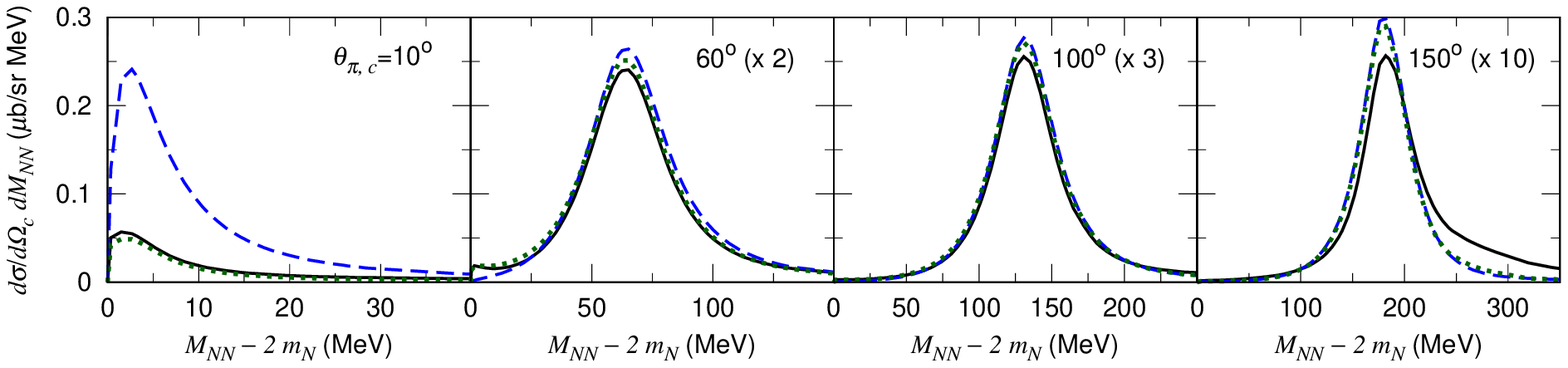}
\caption{The $NN$ invariant mass distribution for
$d(\gamma,\pi^0)pn$ at $E_\gamma = 305$~MeV (upper row)
and $E_\gamma = 705$~MeV (lower row).
The pion emission angle in the CM frame
along with the scaling factor
is indicated in each figure.
Dashed curves: impulse; dotted curves: impulse + $N$-exchange;
solid curves: impulse + $N$-exchange + $\pi$-exchange.
}
 \label{fig:gd-pi0pn-NN}
\end{center}
\end{figure}
\begin{figure}[h]
\begin{center}
\includegraphics[clip,width=1\textwidth]{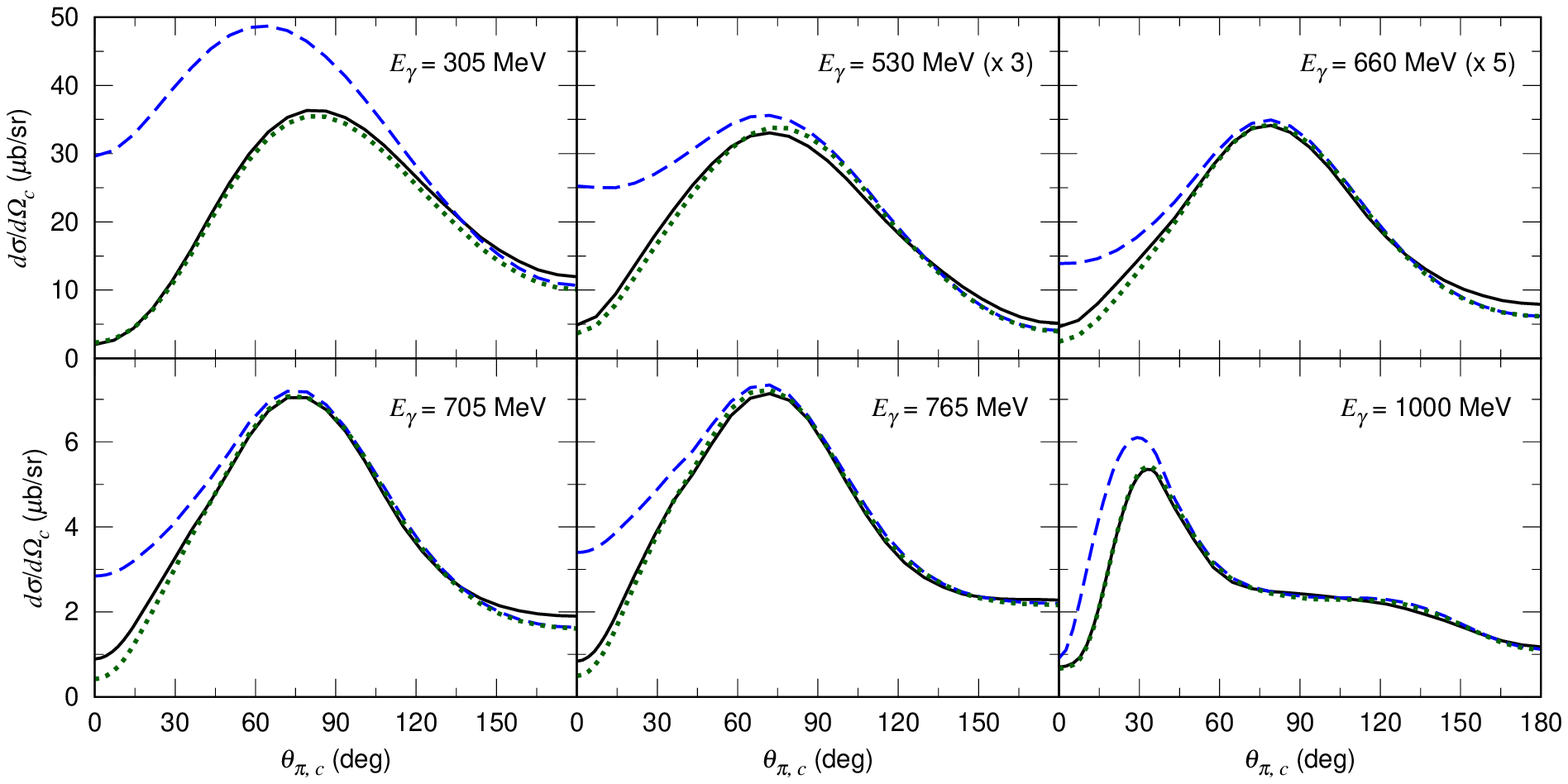}
\caption{
Differential cross sections 
for $d(\gamma,\pi^0)pn$ in the CM frame.
The reaction mechanisms included in the calculations for each of the curves
 are the same as those in Fig.~\ref{fig:gd-pi0pn-NN}.
 }
 \label{fig:gd-pi0pn-dcrst}
\end{center}
\end{figure}

At a higher photon energy of $E_\gamma=705$~MeV
(lower half of Fig.~\ref{fig:gd-pi0pn-NN}), the reduction due to
the $NN$ FSI is similarly dependent on how close
the $M_{NN}$ spectrum peak is to the $NN$ threshold.
We also note that while the $\pi N$ FSI effect is much smaller than the
$NN$ FSI effect overall, it can give a main FSI correction at large pion emission angles;
for example, see the results at $\theta_{\pi,c}\sim 150^\circ$ 
in the lower half of Fig.~\ref{fig:gd-pi0pn-NN}.
Integrating over $M_{NN}$, the FSI effects on the differential cross sections of
$d(\gamma,\pi^0)pn$ are shown in Fig.~\ref{fig:gd-pi0pn-dcrst}. 
The large reduction due to the $NN$ FSI is clear.
Clearly, the large reduction
due to FSI is essential in obtaining a reasonable agreement with the data
in Fig.~\ref{fig:gd-pi0pn-dat}. (Note that the results in
Fig.~\ref{fig:gd-pi0pn-dat} is given in a
different frame chosen in Ref.~\cite{gd-pi0pn_data} to present the data).
This is similar to the previous findings~\cite{arenhover, fix, tara-1}.
The FSI effects at higher
energies are weaker, but still significant in the forward angle, $\theta_{\pi,c}\sim 0^\circ$. 

\begin{figure}[t]
\begin{center}
\includegraphics[clip,width=1\textwidth]{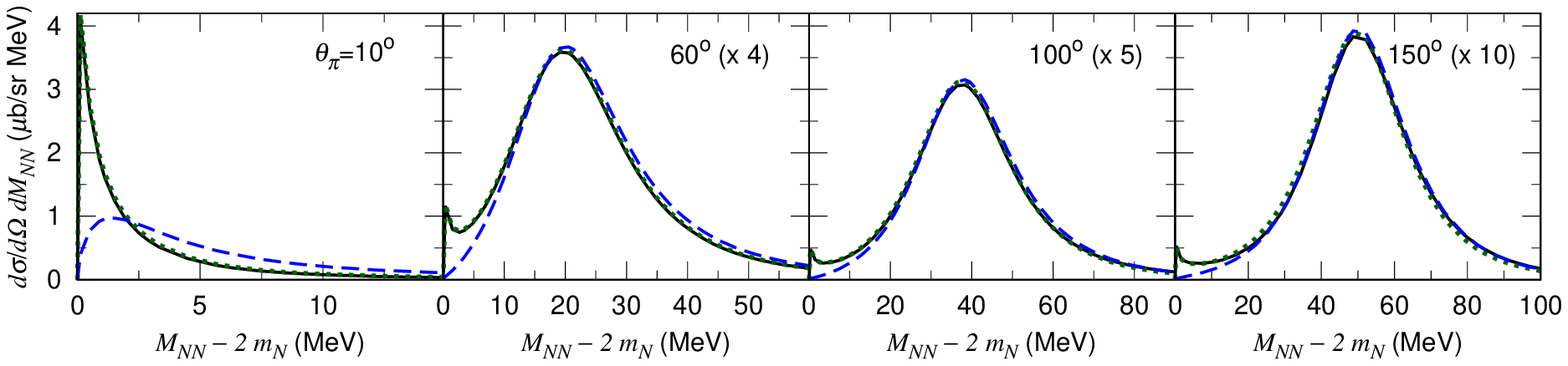}
\includegraphics[clip,width=1\textwidth]{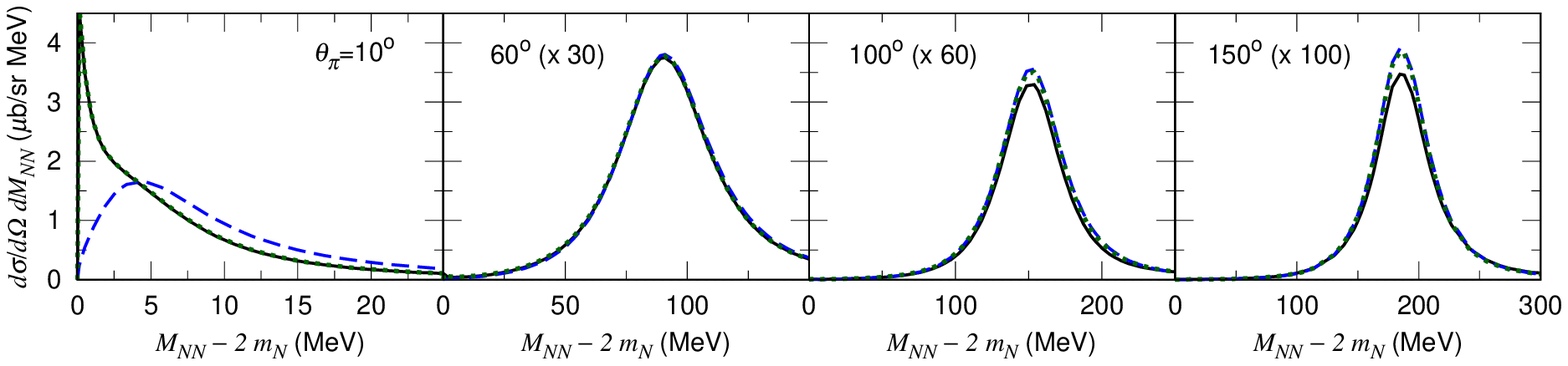}
\caption{The $NN$ invariant mass distribution for
$d(\gamma,\pi^-)pp$ at $E_\gamma = 310$~MeV (upper row)
and $E_\gamma = 700$~MeV (lower row).
The pion emission angles indicated in the figures
are those in the laboratory frame.
The reaction mechanisms included in the calculations for each of the curves
 are the same as those in Fig.~\ref{fig:gd-pi0pn-NN}.
}
 \label{fig:gd-pimpp-NN}
\end{center}
\end{figure}
\begin{figure}[h]
\begin{center}
\includegraphics[clip,width=1\textwidth]{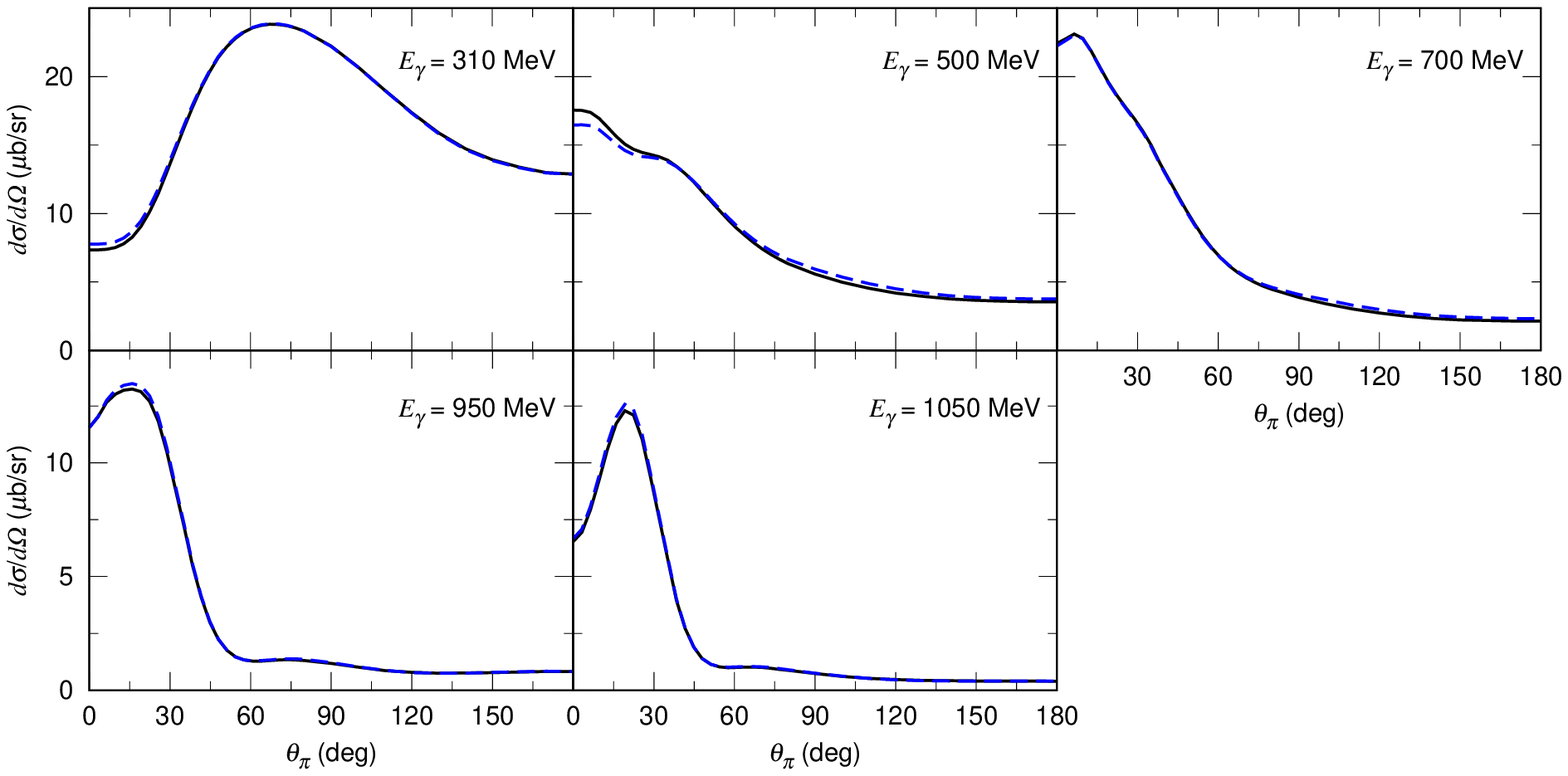}
\caption{Differential cross sections for $d(\gamma,\pi^-)pp$
in the laboratory frame.
The reaction mechanisms included in the calculations for each of the curves
 are the same as those in Fig.~\ref{fig:gd-pi0pn-NN}.
}
 \label{fig:gd-pimpp-dcrst}
\end{center}
\end{figure}
The FSI effects on $d(\gamma,\pi^-)pp$ are shown in Fig.~\ref{fig:gd-pimpp-NN}.
Here, we see that the FSI effect is also clearly visible in the small angle
region where the $M_{NN}$ spectrum peak
is close to the $NN$ threshold and hence the $NN$ FSI going to 
the $^1S_0$ $NN$ partial wave gives a dominant portion of the FSI effect.
Contrary to the $d(\gamma,\pi^0)pn$ case, its interference with the impulse term is 
to enhance the cross section
from the dashed curves to the dotted curves. The $\pi N$ FSI effect is also 
weak here, as can be seen in
the negligible difference between the solid and dotted curves.
As the angle increases,
the $\pi N$ FSI effect is getting discernible but overall
the FSI effect quickly becomes much smaller. 
Therefore, the FSI effect on the differential
cross sections are very weak, as seen in Fig.~\ref{fig:gd-pimpp-dcrst}.

\begin{figure}[h]
\includegraphics[width=1\textwidth]{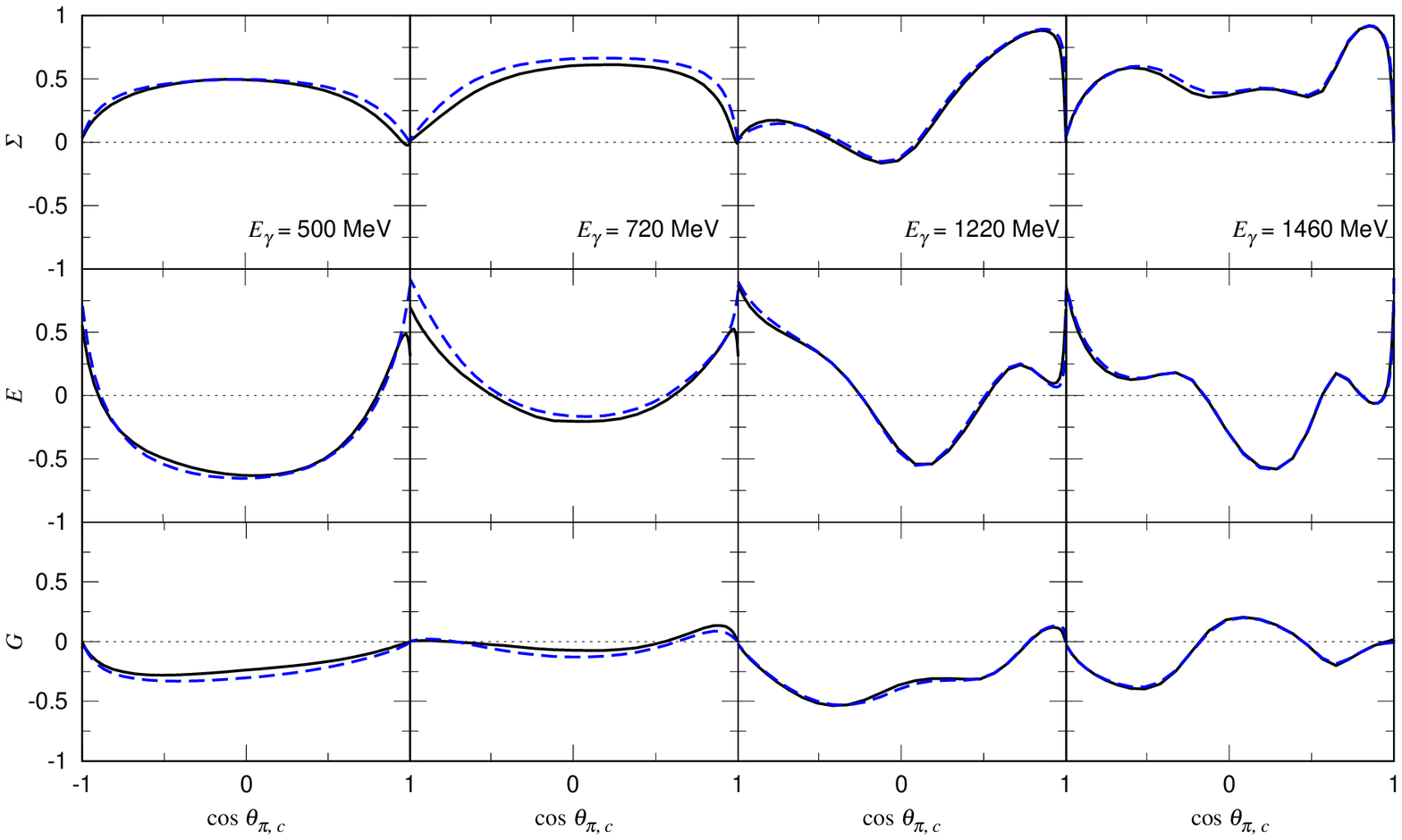}
\caption{
The polarization observables $\Sigma$, $E$, and $G$ for
$d(\gamma,\pi^0)pn$ in the CM frame.
Dashed curves: impulse;
solid curves: impulse + $N$-exchange + $\pi$-exchange.
}
\label{fig:gd-pi0pn-ESG}
\end{figure}
\begin{figure}[h]
\includegraphics[width=1\textwidth]{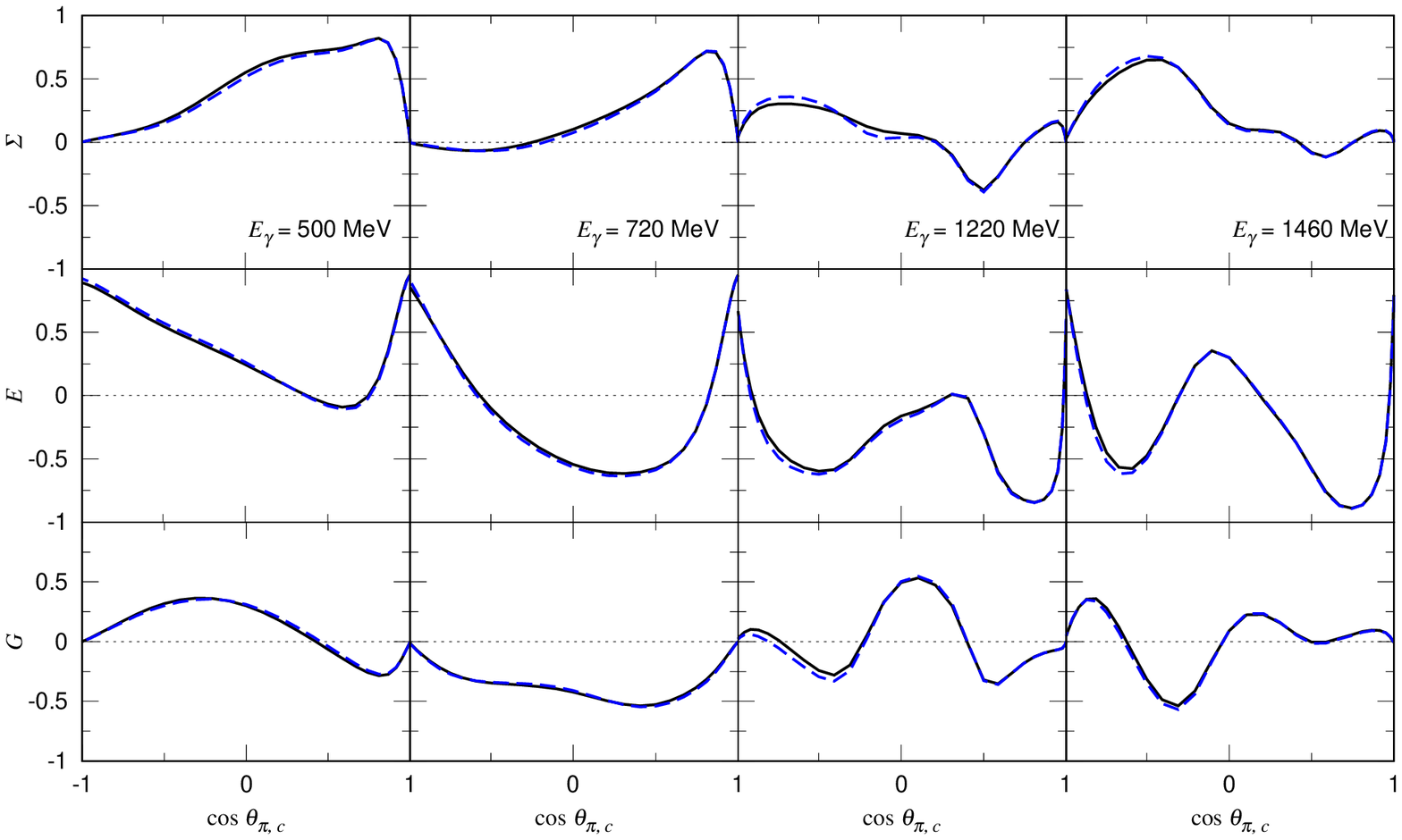}
\caption{
The polarization observables $\Sigma$, $E$, and $G$ for
 $d(\gamma,\pi^-)pp$ in the CM frame.
The other features are the same as those in Fig.~\ref{fig:gd-pi0pn-ESG}.
}
\label{fig:gd-pimpp-ESG}
\end{figure}
We have also investigated the FSI effects on the polarization observables
$\Sigma$, $E$, and $G$ defined in Eqs.~(\ref{eq:sigma})-(\ref{eq:G})
which are of current interests.
The results are shown in Figs.~\ref{fig:gd-pi0pn-ESG} and \ref{fig:gd-pimpp-ESG}.
Clearly, the FSI effects do not play an important role here.
It will be interesting to compare our predictions with the data in near future.

\section{Summary}
\label{sec:summary}

We have applied the ANL-Osaka amplitudes to investigate  
the $d(\gamma,\pi)NN$ in the nucleon resonance region.
Within the multiple scattering formulation, the calculations include 
the impulse term and the final-state interaction terms due to the
pion-exchange  and nucleon-exchange mechanisms.
The predicted differential cross sections are in good agreement with 
most of the available data of $d(\gamma,\pi^0)np$ and
$d(\gamma,\pi^-)pp$ reactions.

We have shown  that
the off-shell effects, calculated from the meson-exchange mechanisms, on the propagations of the
exchanged nucleon and pion  are significant in determining the reaction amplitudes.
The FSI effects on the predicted cross sections
are found to be important at energies near the $\Delta$(1232) resonance,
and are still significant at higher energies. To complete this investigation, we need to
examine the extent to which the predictions presented in this work will be changed when the
$\Delta$-, $\rho$- and $\sigma$-exchange mechanisms, which can also be predicted by using the
ANL-Osaka amplitudes, are included in our calculations.

\begin{acknowledgments}
We thank W.N. Polyzou for his assistance to the use of their relativistic reaction theory.
This work is in part supported by 
Funda\c{c}\~ao de Amparo \`a Pesquisa do Estado de S\~ao Paulo (FAPESP),
Process No.~2016/15618-8, 
by JSPS KAKENHI Grant Numbers 25105010, 16K05354, and 18K03632,
and by the U.S. Department of Energy, Office of Science, Office of Nuclear Physics, Contract 
     No. DE-AC02-06CH11357.
Numerical computations in this work were carried out
with SR16000 at YITP in Kyoto University,
the High Performance Computing system at RCNP in Osaka University,
the National Energy Research Scientific Computing Center, which is
supported by the Office of Science of the U.S. Department of Energy
under Contract No. DE-AC02-05CH11231, 
and the use of the Bebop [or Blues] cluster in the Laboratory Computing
Resource Center at Argonne National Laboratory.

\end{acknowledgments}

\appendix

\section{Ingredients for $\gamma d\to \pi N_1N_2$ matrix elements
in Eqs.~(\ref{eq:amp_imp})-(\ref{eq:amp_MN})}
\label{app1}

The off-energy-shell two-body matrix elements for calculating
Eqs.~(\ref{eq:amp_imp})-(\ref{eq:amp_MN}) are defined
in the $\gamma$-$d$ laboratory frame. 
Here we give formulas for getting these matrix elements from the matrix elements
in the CM frame of $\gamma N$, $\pi N$, and $NN$ systems. 
Our formula are derived from using the procedures of Refs.~\cite{polyzou, polyzou-1}
within the instant form of relativistic quantum mechanics~\cite{kei-pol}. The details of this approach
 can be found in these references. Here we only give the formulas used in our numerical calculations. 

\subsection{$\gamma N\rightarrow \pi N'$ matrix elements}
The  off-shell $\gamma N\to\pi N'$ matrix element $\bra{\pi N'}t_{\pi N,\gamma N}(W) \ket{\gamma N}$
in Eqs.~(\ref{eq:amp_imp})-(\ref{eq:amp_MN}),
is related to those evaluated in the CM frame of $\pi N'$  by
\begin{eqnarray}
&&\bra{\pi(\bm{k},t_\pi)\, N'(\bm{p}',s',t')} t_{\pi N, \gamma N}(W)
\ket{\gamma(\bm{q},\lambda)\, N(\bm{p} ,s,t)}
\nonumber \\
&=&\bra{\pi(\bm{k},t_\pi)\, N'(\bm{p}',s',t')} \epsilon^\mu_\lambda(q)[j(W)]_\mu
\ket{ N(\bm{p} ,s,t)}
\nonumber \\
&=&
\sqrt{|\bm{q}_{\bar c}|E_\pi(\bm{k}_{\bar c})E_N(-\bm{q}_{\bar c})E_N(-\bm{k}_{\bar c})\over  |\bm{q}|E_\pi(\bm{k})E_N(\bm{p})E_N(\bm{p}')}
\sum_{\mu\nu}\epsilon^\mu_\lambda(q)\Lambda_\mu^\nu(p_t)\sum_{s_{\bar c},s'_{\bar c}}
R^*_{s'_{\bar c},s'}(p',p_t)
R_{s_{\bar c},s}(p,p_t)
\nonumber \\
&\times&
\bra{\pi(\bm{k}_{\bar c},t_\pi) N'(-\bm{k}_{\bar c},s'_{\bar c},t')} [j_{\bar c}(W)]_\nu
\ket{N(-\bm{q}_{\bar c},s_{\bar c},t)} \ ,
\label{eq:amp_imp_lor-a}
\end{eqnarray}
where the suffixes '$\bar c$' indicate quantities in the CM system of $\pi N'$,
$\epsilon^\mu_\lambda(q)$ is the photon polarization vector, $j_\mu(W)$ is the current operator,
$p_t =p'+k$ is a four-momentum defined by
$\bm{p}_t=\bm{p}'\!+\!\bm{k}$ and
$p_t^0=E_{N}(\bm{p}')\!+\!E_\pi(\bm{k})$.
The CM matrix elements of the current operator, 
$\bra{\pi(\bm{k}_{\bar c},t_\pi) N'(-\bm{k}_{\bar c},s'_{\bar c},t')} [j_{\bar c}(W)]_\mu]
\ket{N(-\bm{q}_{\bar c},s_{\bar c},t)}$, from the ANL-Osaka model 
are calculated following the procedure 
detailed in Appendix~D of Ref.~\cite{knls13},
and their normalization is specified by Eq.~(D2) of the reference.

In Eq.~(\ref{eq:amp_imp_lor-a}), the quantity $\Lambda^\nu_\mu(p_t)$ boosts 
the CM momenta $a_{\bar c}=(k^0_{\bar c},\bm{k}_{\bar c})$ 
[$ (q^0_{\bar c},\bm{q}_{\bar c})$] to the Lab momenta 
$a_L=(k^0, \bm{k})$ [$(q^0,\bm{q})$] by the  Lorentz transformation,
\begin{eqnarray}
a_{L}^0 &=&\sum_{\nu}\Lambda^0_\nu(p_t) a^\nu_{\bar c}
= {a_{\bar c}^0\, p_t^0 + \bm{p}_t\cdot\bm{a}_{\bar c} \over m_t}\ ,
\nonumber \\
a_{L}^i &=& \sum_{\nu}\Lambda^i_\nu(p_t) a^\nu_{\bar c}
= a_{\bar c}^i + p_t^i\left[{ \bm{p}_t\cdot\bm{a}_{\bar c}\over m_t (m_t+p_t^0)}
+{ {a}^0_{\bar c}\over m_t}\right] \ ,
\label{eq:lorentz}
\end{eqnarray}
where the index $i=1,2,3$ is a spatial component and $m_t\equiv\sqrt{p_t \cdot p_t}$.

The spin rotation matrix $R_{s_{\bar c},s}(p,p_t)$ in Eq.~(\ref{eq:amp_imp_lor-a}) 
is given~\cite{polyzou,polyzou-1} explicitly as:
\begin{eqnarray}
R_{s_{\bar c},s}(p,p_t) &=& 
\bra{s_{\bar c}} B^{-1}(p_{\bar c}/m_N)B^{-1}(p_t/m_t)B(p/m_N) \ket{s} \ ,
\label{eq:spin-rot}
\end{eqnarray}
where $\ket{s_{(\bar c)}}$ is the spin state of the nucleon,
$p_{\bar c}$ is obtained from the nucleon momentum $p$ in the laboratory frame by
the Lorentz transformation of Eq.~(\ref{eq:lorentz}),
and
\begin{eqnarray}
B (p/m) &=& \frac{1}{\sqrt{2m\,(p^0+m)}} ((p^0+m)\unittwo +
\bm{p}\cdot \bm{\sigma}) \ , \nonumber \\
B ^{-1}(p/m) &=& \frac{1}{\sqrt{2m\,(p^0+m)}} ((p^0+m)\unittwo -
\bm{p}\cdot \bm{\sigma}) \ ,
\label{eq:lb-tx}
\end{eqnarray}
where $\bm{\sigma}$ is the Pauli operator and $\unittwo$ is the unit matrix.

\subsection{$\pi' N'\to \pi N$ matrix element}
\label{app2}

The half-off-shell 	$\pi' N'\to \pi N$ scattering amplitudes in the laboratory frame,
appearing in Eq.~(\ref{eq:amp_MN}),
are related to those in the $\pi N$ CM frame by
\begin{eqnarray}
&&\bra{\pi(\bm{k},t_\pi) N(\bm{p},s,t)}t_{\pi N}(M_{\pi N})
\ket{\pi'(\bm{k}',t_\pi') N'(\bm{p}',s',t')}
\nonumber \\
&=&
\sqrt{
E_{\pi}(\bm{k}_{\bar c})
E_{N}(-\bm{k}_{\bar c})
E_{\pi}(\bm{k}'_{\bar c})
E_N(-\bm{k}'_{\bar c})
\over
E_{\pi}(\bm{k})
E_{N}(\bm{p})
E_{\pi}(\bm{k}')
E_{N}(\bm{p}')
}
\sum_{s_{\bar c},s'_{\bar c}}
R^*_{s_{\bar c},s}(p,p_t) R_{s'_{\bar c},s'}(p',p_t)
\nonumber \\
&\times&
\bra{\pi(\bm{k}_{\bar c},t_\pi) N(-\bm{k}_{\bar c},s_{\bar c},t)}
t^{\bar c}_{\pi N}(M_{\pi N})\ket{\pi'(\bm{k}'_{\bar c},t'_\pi) N'(-\bm{k}'_{\bar c},s'_{\bar c},t')}
\ ,
\label{eq:amp_MN_lor}
\end{eqnarray}
where the momentum $\bm{k}_{\bar c}$ in the $\pi N$ CM frame is related to
${\bm{k}}$ in the laboratory frame by the
Lorentz transformation  Eq.~(\ref{eq:lorentz}). Here we choose 
$\bm{p}_t= \bm{k}+\bm{p}$ and $p_t^0= E_\pi(\bm{k})+E_{N}(\bm{p})$.
Within our formulation
where an energy of a particle $x$ is always related to its off-shell
momentum ($\bm{p}_x$) by $E_x=\sqrt{|\bm{p}_x|^2+m_x^2}$, 
the same Lorentz transformation does not bring the off-shell $\pi'N'$
system in the laboratory frame into their CM frame.
Therefore, we define the $\pi'N'$ relative momentum in the $\pi N$ CM
frame by 
$\bm{k}'_{\bar c}= (m_N \tilde{\bm{k}}'_{\bar c} - m_\pi \tilde{\bm{p}}'_{\bar c} )/(m_N + m_\pi)$,
where $\tilde{\bm{k}}'_{\bar c}$ and $\tilde{\bm{p}}'_{\bar c}$ are 
related to $\bm{k}'$ and $\bm{p}'$ by the same Lorentz transformation Eq.~(\ref{eq:lorentz}).
The $\pi N$ scattering amplitudes in the $\pi N$ CM frame,
$\bra{\pi(\bm{k}_{\bar c},t_\pi) N(-\bm{k}_{\bar c},s_{\bar c},t)}
t^{\bar c}_{\pi N}(M_{\pi N})\ket{\pi'(\bm{k}'_{\bar c},t'_\pi) N'(-\bm{k}'_{\bar c},s'_{\bar c},t')}$,
are calculated following the procedure 
detailed in Appendix~C of Ref.~\cite{knls13},
and their normalization is specified by Eq.~(C7) of the reference.

\subsection{$N'_1N'_2\to N_1N_2$ matrix element}
\label{app3}

The half off-shell $N'_1N'_2\to N_1N_2$ matrix elements in the laboratory frame,
appearing in Eq.~(\ref{eq:amp_NN}),
are related to those in the $N_1N_2$ CM frame by
\begin{eqnarray}
&&\bra{N_1(\bm{p}_1,s_1,t_1)N_2(\bm{p}_2,s_2,t_2)}
T_{NN}(M_{N_1N_2})
\ket{N'_1(\bm{p}'_1,s'_1,t_1)N'_2(\bm{p}'_2,s'_2,t_2)}
\nonumber \\ 
&=& 
\sqrt{
E^2_N(\bm{p}_{\bar c})
E^2_N(\bm{p}_{\bar c}')
\over
E_{N}(\bm{p}_1)
E_{N}(\bm{p}_2)
E_{N}(\bm{p}'_1)
E_{N}(\bm{p}'_2)
}
\sum_{s_{1\bar c},s_{2\bar c},s'_{1\bar c},s'_{2\bar c}}
R^*_{s_{1\bar c},s_1}(p_1,p_t)
R^*_{s_{2\bar c},s_2}(p_2,p_t)
R_{s'_{1\bar c},s'_1}(p'_1,p_t)
R_{s'_{2\bar c},s'_2}(p'_2,p_t)
\nonumber \\ 
&\times&
\bra{N_1(\bm{p}_{\bar c},s_{1\bar c},t_1)N_2(-\bm{p}_{\bar c},s_{2\bar c},t_2)}
T^{\bar c}_{NN}(M_{N_1N_2})
\ket{N'_1(\bm{p}'_{\bar c},s'_{1\bar c},t_1)N'_2(-\bm{p}'_{\bar c},s'_{2\bar c},t_2)}
\ ,
\label{eq:t_NN_lor}
\end{eqnarray}
with $p_t$ being
$\bm{p}_t= \bm{p}_1+\bm{p}_2$ and $p_t^0= E_N(\bm{p}_1)+E_{N}(\bm{p}_2)$.
Regarding the $NN$ relative momenta, we use
the non-relativistic relation 
$\bm{p}_{\bar c}=(\bm{p}_1-\bm{p}_2)/2$ and
$\bm{p}'_{\bar c}=(\bm{p}'_1-\bm{p}'_2)/2$.
The $NN$ $T$-matrix $\langle T^{\bar c}_{NN}\rangle$
is expanded in terms of partial waves as
\begin{eqnarray}
&&
\bra{N_1(\bm{p}_{\bar c},s_{1\bar c},t_1)N_2(-\bm{p}_{\bar c},s_{2\bar c},t_2)}
T^{\bar c}_{NN}(M_{N_1N_2})
\ket{N'_1(\bm{p}'_{\bar c},s'_{1\bar c},t_1)N'_2(-\bm{p}'_{\bar c},s'_{2\bar c},t_2)}
\nonumber\\
&=& 
\sum_{JLL'ST}
\left({1-(-1)^{L+S+T}} \over 2^{|t_1+t_2|}\right)
\sum_{S^zS^{\prime z}L^zL^{\prime z}J^zT^z}
t^{JLL'ST}_{NN} (|\bm{p}_{\bar c}|,|\bm{p}'_{\bar c}|;M_{N_1N_2})
\nonumber\\
&\times&
(1/2,t_1,1/2,t_2|TT^z)(1/2,s_{1\bar c},1/2,s_{2\bar c}|SS^z)
(LL^zSS^z|JJ^z) Y_{LL^z}(\hat{p}_{\bar c})
\nonumber\\
&\times&
(1/2,t_1,1/2,t_2|TT^z)(1/2,s'_{1\bar c},1/2,s'_{2\bar c}|SS^{\prime z}) 
(L'L^{\prime z}SS^{\prime z}|JJ^z)
Y^*_{L'L^{\prime z}}(\hat{p}'_{\bar c}) \ ,
\label{eq:t_NN}
\end{eqnarray}
where 
$t^{JLL'ST}_{NN}$
is a partial wave $NN$ amplitude 
characterized by the quantum numbers $JLL'ST$ meaning of which are
self-evident in the above equation.
On- and off-shell partial wave $NN$ amplitudes are
obtained by solving the following
Lippmann-Schwinger equation:
\begin{eqnarray}
t^{JLL'ST}_{NN} (p,p';M_{N_1N_2})
=
v^{JLL'ST}_{NN} (p,p')
+
\sum_{L''}
\int^\infty_0\!\!\! q^2 dq 
{v^{JLL''ST}_{NN} (p,q)\,
t^{JL''L'ST}_{NN} (q,p';M_{N_1N_2})
\over
M_{N_1N_2}-2m_N - {q^2\over m_N} + i\epsilon
} ,\nonumber \\
\label{eq:LS}
\end{eqnarray}
where we use the CD-Bonn potential~\cite{cdbonn}
for 
$v^{JLL'ST}_{NN} (p',p)$.

\subsection{Deuteron wave function}
\label{app4}

The deuteron wave function in its rest frame, 
appearing in Eqs.~(\ref{eq:amp_imp})-(\ref{eq:amp_MN}),
is more explicitly written as
\begin{eqnarray}
\inp{N_1(\bm{p},s_1,t_1)\, N_2 (-\bm{p},s_2,t_2)}{\Psi_d(s_d)}
&=& 
\left[
{\delta_{s,s_d}\over\sqrt{4\pi}}u_s(p)
+(2,s_d\!-\!s,1,s|1 s_d) u_d(p)Y_{2,s_d-s}(\hat{p})
\right]
\nonumber \\
& \times& (1/2, s_1, 1/2, s_2 | 1 s)
 (1/2, t_1, 1/2, t_2 | 0 0) \ ,
\label{eq:wf_deu}
\end{eqnarray}
with $s=s_1+s_2$.
The radial $s$- and $d$-wave functions, denoted by $u_s$ and $u_d$
respectively, are normalized as 
$\int_0^\infty dp p^2 (|u_s(p)|^2 + |u_d(p)|^2)=1$.
We use the CD-Bonn potential~\cite{cdbonn} to generate the deuteron wave function.
The nucleon momentum distribution in the deuteron is given by
\begin{eqnarray}
\rho_d(p)=|u_s(p)|^2 + |u_d(p)|^2 \ .
\label{eq:mom_deu}
\end{eqnarray}



\end{document}